\documentclass[a4paper,11pt]{article}
\pdfoutput=1
\usepackage{jheppub}

\usepackage{amssymb,amsmath}
\usepackage[normalem]{ulem}
\usepackage[utf8x]{inputenc}
\usepackage{slashed}
\usepackage{graphicx}

\usepackage{csquotes} 
\usepackage{comment}
\usepackage{mathrsfs}
\usepackage{float}

\newcommand{\talpha}{\tilde{\alpha}}
\newcommand{\tbeta}{\tilde{\beta}}
\newcommand{\tgamma}{\tilde{\gamma}}

\def\lsim{\mathrel{\rlap{\lower4pt\hbox{\hskip1pt$\sim$}}
    \raise1pt\hbox{$<$}}}         
\def\gsim{\mathrel{\rlap{\lower4pt\hbox{\hskip1pt$\sim$}}
    \raise1pt\hbox{$>$}}}         

\newcommand{\drawsquare}[2]{\hbox{%
\rule{#2pt}{#1pt}\hskip-#2pt
\rule{#1pt}{#2pt}\hskip-#1pt
\rule[#1pt]{#1pt}{#2pt}}\rule[#1pt]{#2pt}{#2pt}\hskip-#2pt
\rule{#2pt}{#1pt}}
\newcommand{\fund}{\raisebox{-.5pt}{\drawsquare{6.5}{0.4}}}
\newcommand{\Ysymm}{\raisebox{-.5pt}{\drawsquare{6.5}{0.4}}\hskip-0.4pt%
        \raisebox{-.5pt}{\drawsquare{6.5}{0.4}}}
\newcommand{\Yasymm}{\raisebox{-3.5pt}{\drawsquare{6.5}{0.4}}\hskip-6.9pt%
        \raisebox{3pt}{\drawsquare{6.5}{0.4}}}
\newcommand{\antifund}{\overline{\fund}}

\numberwithin{equation}{section}

\preprint{
\begin{minipage}{5cm}
\small
\flushright
KEK-TH-2530\\KYUSHU-HET-263
\end{minipage}}

\title{Flavor, CP and Metaplectic Modular Symmetries in Type IIB Chiral Flux Vacua}

\author{Keiya Ishiguro$^{1}$,} 
\author{Takafumi Kai$^{2}$,} 
\author{Hiroshi Okada$^{3,4}$, and} 
\author{Hajime Otsuka$^{2}$} 
\affiliation{
$^1$ Graduate University for Advanced Studies (Sokendai), 1-1 Oho, Tsukuba, Ibaraki 305-0801, Japan\\
$^2$ Department of Physics, Kyushu University, 744 Motooka, Nishi-ku, Fukuoka 819-0395, Japan\\
$^3$Asia Pacific Center for Theoretical Physics (APCTP), Pohang 37673, Republic of Korea \\
$^4$Department of Physics, Pohang University of Science and Technology, Pohang 37673, Republic of Korea \\
}
\emailAdd{ishigu@post.kek.jp}
\emailAdd{kai.takafumi.606@s.kyushu-u.ac.jp}
\emailAdd{hiroshi.okada@apctp.org}
\emailAdd{otsuka.hajime@phys.kyushu-u.ac.jp}

\abstract{
We examine symmetries of chiral four-dimensional vacua of Type IIB flux compactifications with vanishing superpotential $W=0$. 
We find that the ${\cal N}=1$ supersymmetric MSSM-like and Pati-Salam vacua possess enhanced discrete symmetries in the effective action below the mass scale of stabilized complex structure moduli and dilaton. 
Furthermore, a generation number of quarks/leptons is small on these vacua where the flavor, CP and metaplectic modular symmetries are described in the framework of eclectic flavor symmetry. 
}
\makeatletter
\gdef\@fpheader{}
\makeatother

\begin{document}

\maketitle

\section{Introduction}
\label{sec:intro}

The string theory predicts a huge number of low-energy effective field theories, 
the so-called string theory landscape. 
In particular, background fluxes in extra-dimensional spaces lead to 
the rich and attractive vacuum structure of the string landscape, which 
will be quantified by a statistical study \cite{Ashok:2003gk,Douglas:2003um,Denef:2004ze} as well as the swampland program \cite{Vafa:2005ui,ArkaniHamed:2006dz,Ooguri:2006in}\footnote{See for a review, e.g., Ref. \cite{Palti:2019pca}.}. 
It is known that the statistical study of Type IIB flux vacua is a powerful approach to address 
the vacuum distribution and selection rules on the moduli spaces. 

\medskip

In Type IIB flux compactifications on $T^6/(\mathbb{Z}_2 \times \mathbb{Z}_2^\prime)$ orientifolds, 
the distribution of complex structure moduli fields was known to be clustered at fixed points of $SL(2,\mathbb{Z})$ modular symmetry of the torus \cite{DeWolfe:2004ns,Ishiguro:2020tmo}, where the fixed points in the $SL(2,\mathbb{Z})$ 
moduli space correspond to $\tau = i, \omega, i\infty$ with $\omega = \frac{-1+\sqrt{3}i}{2}$, each with 
enhanced symmetries. 
Remarkably, probabilities of moduli values are peaked at a $\mathbb{Z}_3$ fixed point $\tau = \omega$, indicating that discrete $\mathbb{Z}_3$ symmetry remains in the low-energy effective action of moduli fields \cite{Ishiguro:2020tmo}. 
Such a novel feature about the distribution of flux vacua was explored in this simple toroidal orientifold 
but it is expected to appear in a more generic Calabi-Yau moduli space with symplectic modular symmetry.

\medskip

In this paper, we further examine semi-realistic four-dimensional (4D) vacua with Standard Model (SM) spectra. 
Since a generation number of fermions is determined by background fluxes on magnetized D-branes, 
the generation number and three-form fluxes stabilizing moduli fields will be correlated through tadpole cancellation conditions of D-branes. 
It is interesting to reveal how much 3-generation models are distributed in the flux landscape. 
Furthermore, the flavor symmetries of quarks and leptons will also be related to the modular symmetry 
of the torus, because moduli-dependent Yukawa couplings transform under the moduli symmetry \cite{Cremades:2004wa}. 
For illustrative purpose, we deal with the simple $T^6/(\mathbb{Z}_2 \times \mathbb{Z}_2^\prime)$ orientifolds.
By analyzing physically-distinct configurations of background fluxes leading to vanishing superpotential $W=0$, 
we find that the generation number of quarks and leptons is restricted to be small due to the 
tadpole cancellation condition. 
Furthermore, 
flavor symmetry, CP, and modular symmetry in semi-realistic 4D vacua are uniformly described in the context of 
eclectic flavor symmetry \cite{Nilles:2020nnc,Nilles:2020kgo} as developed in both the top-down and bottom-up approaches \cite{Nilles:2020nnc,Nilles:2020kgo,Baur:2020jwc,Baur:2022hma,Nilles:2020gvu,DNilles:2020gvu}. 

\medskip

This paper is organized as follows. 
In Sec. \ref{sec:2}, we first review the flux compactifications on $T^6/(\mathbb{Z}_2 \times \mathbb{Z}_2^\prime)$. Next, we incorporate specific magnetized D-brane models without and with a discrete $B$ field in Secs. \ref{sec:generation1} and \ref{sec:generation2}, respectively. It turned out that the string landscape leads to the small generation number of quarks and leptons. 
In Sec. \ref{sec:eclectic}, we begin with the metaplectic modular symmetry in Sec. \ref{sec:meta}, which can be realized in $T^2$ and $T^2/\mathbb{Z}_2$ orbifold with magnetic fluxes as discussed in Secs. \ref{sec:T2} and \ref{sec:T2Z2}, respectively. 
The CP transformation will be unified in the context of generalized modular symmetry in Sec. \ref{sec:CP}. Finally, we discuss the unification of flavor, CP, and modular symmetries in Type IIB chiral 4D flux vacua in Sec. \ref{sec:10D}. 
Sec. \ref{sec:con} is devoted to the conclusion.

\section{Moduli distributions in Type IIB flux vacua with SM spectra}
\label{sec:2}

In Sec. \ref{sec:moduli}, we first review the vacuum structure of Type IIB flux compactification on $T^6/(\mathbb{Z}_2 \times \mathbb{Z}_2^\prime)$ orientifolds. 
Next, we introduce semi-realistic magnetized D-brane models in Type IIB 
flux vacua, taking into account the tadpole cancellation conditions in Secs. \ref{sec:generation1} and \ref{sec:generation2}. 
It is found that the generation number of quarks and leptons is restricted to be small due to the tadpole cancellation condition.

\subsection{Flux compactifications on $T^6/(\mathbb{Z}_2 \times \mathbb{Z}_2^\prime)$ orientifolds}
\label{sec:moduli}

We begin with the Type IIB flux compactifications on $T^6/(\mathbb{Z}_2 \times \mathbb{Z}_2^\prime)$ orientifolds without discrete torsion, following the notation of Ref. \cite{Blumenhagen:2006ci}. 
The complex coordinates of $T^6= (T^2)_1 \times (T^2)_2 \times (T^2)_3$ are defined as $z_i = x_i +\tau_i y_i$ with $i=1,2,3$, subject to the following $\mathbb{Z}_2$ orbifoldings:
\begin{align}
    \theta &: (z_1, z_2, z_3) \rightarrow (-z_1, -z_2, z_3),
    \nonumber\\
    \theta^\prime &: (z_1, z_2, z_3) \rightarrow (z_1, -z_2, -z_3).
    \label{eq:projections}
\end{align}
Furthermore, the orientifold projection acts $z_i$ as
\begin{align}
    {\cal R} : (z_1, z_2, z_3) \rightarrow (-z_1, -z_2, -z_3),
\end{align}
in addition to the world-sheet parity $\Omega$. 
It results in 64 O3-planes, 4O$7_1$-, 4O$7_2$-, 4O$7_3$-planes, located at a fixed point of ${\cal R}$ 
and fixed locus of ${\cal R}\theta'$, ${\cal R}\theta\theta'$, ${\cal R}\theta$, respectively.

\medskip

On the toroidal ambient space, i.e., $T^6$, the following three forms are invariant under $\mathbb{Z}_2 \times \mathbb{Z}_2^\prime$ symmetry:
\begin{align}
    \alpha_0 &= dx_1 \wedge dx_2 \wedge dx_3, 
    \qquad
    \beta^0 = dy_1 \wedge dy_2 \wedge dy_3,     
    \nonumber\\
    \alpha_1 &= dy_1 \wedge dx_2 \wedge dx_3, 
    \qquad
    \beta^1 = -dx_1 \wedge dy_2 \wedge dy_3,     
    \nonumber\\
    \alpha_2 &= dx_1 \wedge dy_2 \wedge dx_3, 
    \qquad
    \beta^2 = -dy_1 \wedge dx_2 \wedge dy_3,     
    \nonumber\\
    \alpha_3 &= dx_1 \wedge dy_2 \wedge dx_3, 
    \qquad
    \beta^3 = -dy_1 \wedge dy_2 \wedge dx_3. 
    \label{eq:threeforms}
\end{align}
with $\int_{T^6} \alpha_I \wedge \beta^J = \delta_I^J$. To define the complex structure moduli $\tau_i$, we expand the holomorphic three-form on 
the above bases:
\begin{align}
    \Omega_3 = dz_1 \wedge dz_2 \wedge dz_3 = \sum_{I=0}^{3} \left( X^I \alpha_I - F_I \beta^I \right),
\end{align}
where the homogeneous coordinates $X^I$ and the derivatives $F_I=\partial_{X^I} F$ of the prepotential $F$ are defined as
\begin{align}
    X^I = \int_{A^I} \Omega_3,\qquad F_I = \int_{B_I}\Omega_3.
\end{align}
These explicit forms are now given by
\begin{align}
    X^0 &= 1,\qquad
    F_0  = -\tau_1\tau_2\tau_3,
    \nonumber\\
    X^1 &= \tau_1,\qquad
    F_1  = \tau_2\tau_3,
    \nonumber\\
    X^2 &= \tau_2,\qquad
    F_2  = \tau_1\tau_3,
    \nonumber\\
    X^3 &= \tau_3,\qquad
    F_3  = \tau_1\tau_2.
\end{align}
Note that the three-cycles $\{A^I, B_I\}$ correspond to the Poincare-dual basis of the three-forms (\ref{eq:threeforms}), satisfying $A^I \cap A^J = B_I  \cap B_J = 0$ and $A^I \cap B_J = \delta^I_J$ with $I, J = 0,...,3$. 
The $\Pi_{i=1}^3 SL(2,\mathbb{Z})_i$ modular symmetries of the factorizable torus $(T^2)_i$ can be seen on these coordinates. As discussed in Ref. \cite{Ishiguro:2021ccl}, two generators of $\Pi_{i=1}^3 SL(2,\mathbb{Z})_i$: 
\begin{align}
    S_{(i)} : \tau_i \rightarrow -1/\tau_i,
    \quad
    T_{(i)} : \tau_i \rightarrow \tau_i+1,    
\end{align}
are embedded into $8\times 8$ matrices, e.g., 
 \begin{align}
 \begin{split}
S_{(1)}&=
\begin{pmatrix}
0 & -1 & 0 & 0 & 0 & 0 & 0 & 0\\
1 & 0 & 0 & 0 & 0 & 0 & 0 & 0\\
0 & 0 & 0 & 0 & 0 & 0 & 0 & -1\\
0 & 0 & 0 & 0 & 0 & 0 & -1 & 0\\
0 & 0 & 0 & 0 & 0 & -1 & 0 & 0\\
0 & 0 & 0 & 0 & 1 & 0 & 0 & 0\\
0 & 0 & 0 & 1 & 0 & 0 & 0 & 0\\
0 & 0 & 1 & 0 & 0 & 0 & 0 & 0
\end{pmatrix}
, \quad 
T_{(1)}=
\begin{pmatrix}
1 & 0 & 0 & 0 & 0 & 0 & 0 & 0\\
1 & 1 & 0 & 0 & 0 & 0 & 0 & 0\\
0 & 0 & 1 & 0 & 0 & 0 & 0 & 0\\
0 & 0 & 0 & 1 & 0 & 0 & 0 & 0\\
0 & 0 & 0 & 0 & 1 & -1 & 0 & 0\\
0 & 0 & 0 & 0 & 0 & 1 & 0 & 0\\
0 & 0 & 0 & 1 & 0 & 0 & 1 & 0\\
0 & 0 & 1 & 0 & 0 & 0 & 0 & 1
\end{pmatrix}
.
\label{eq:S1T1}
\end{split}
\end{align}
The other generators are also constructed by flipping the corresponding moduli fields. 
Thus, the modular groups are the subgroup of $Sp(8,\mathbb{Z})$, which is the symplectic modular symmetry of the complex structure moduli space in the homogeneous coordinates.

\medskip

It was known that the 4D kinetic terms of the closed string moduli, i.e., three complex structure moduli $\tau_i$, the axio-dilaton $S$ and three K\"ahler moduli $T_i$ are derived from the following K\"ahler potential in units of the reduced Planck mass $M_{\rm Pl}=1$:
\begin{align}
    K = -\ln (i(\tau_1 - \Bar{\tau}_1)(\tau_2 - \Bar{\tau}_2)(\tau_3 - \Bar{\tau}_3))-\ln (i(\Bar{S}-S)) -2\ln {\cal V}(T,\Bar{T}),
\end{align}
where ${\cal V}$ denotes the torus volume in units of the string length $l_s = 2\pi \sqrt{\alpha^\prime}$. 
The moduli superpotential is induced by background three-form fluxes in Type IIB string theory. 
Throughout this paper, we focused on the stabilization of complex structure moduli and axio-dilaton. 
Let us introduce the background Ramond-Ramond (RR) $F_3$ and Neveu-Schwarz three forms $H_3$ as follows:
\begin{align}
\frac{1}{l_s^2}F_3 &= a^0 \alpha_0 +a^{i}\alpha_{i} +b_{i}\beta^{i} +b_0 \beta^0,
\nonumber\\
\frac{1}{l_s^2}H_3 &= c^0 \alpha_0 +c^{i}\alpha_{i} +d_{i}\beta^{i} +d_0 \beta^0,
\label{eq:F3H3}
\end{align}
where $\{a^{0,1,2,3},b_{0,1,2,3},c^{0,1,2,3},d_{0,1,2,3}\}$ correspond to the integral flux quanta. 
They lead to the flux-induced superpotential in the 4D effective action~\cite{Gukov:1999ya}:
\begin{align}
    W &= \frac{1}{l_s^2}\int G_3 \wedge \Omega
    \nonumber\\
    &= a^0 \tau_1\tau_2\tau_3  +c^1 S \tau_2\tau_3
    +c^2 S\tau_1\tau_3 +c^3 S\tau_1\tau_2
 - \sum_{i=1}^3 b_i \tau_i + d_0 S
 \nonumber\\
& -c^0 S\tau_1\tau_2\tau_3 - a^1 \tau_2\tau_3
    - a^2\tau_1\tau_3 - a^3\tau_1\tau_2
+ \sum_{i=1}^3 d_i S \tau_i - b_0. 
    \label{eq:W}
\end{align}
In Ref. \cite{Ishiguro:2020nuf}, the moduli stabilization was performed in the isotropic regime, namely
\begin{align}
\tau \equiv \tau_1=\tau_2 =\tau_3,
\label{eq:isotropic}
\end{align}
with overall flux quanta:
\begin{align}
a\equiv a^1=a^2=a^3,\quad b\equiv b_1=b_2=b_3,\quad
c\equiv c^1=c^2=c^3,\quad d\equiv d_1=d_2=d_3.
\end{align}
The moduli vacuum expectation values (VEVs) are given by
\begin{align}
    \langle S \rangle &=  \frac{ r \tau + s}{u \tau + v}\,,
    \label{eq:Svev}
\end{align}
for the axio-dilaton and
\begin{align}
\begin{split}
    \langle \tau \rangle &= \frac{ - m + \sqrt{m^2 - 4 l n}}{2 l} \quad (l, n > 0)\,,
    \nonumber\\
    \langle \tau \rangle &= \frac{ - m - \sqrt{m^2 - 4 l n}}{2 l} \quad (l, n < 0)\,,
\end{split}
    \label{eq:tauvev}
\end{align}
for the overall complex structure modulus, respectively. 
Here, we redefine the flux quanta
\begin{alignat}{4}
    r l &=  a^0,~ &  r m + s l &=  -3 a,~ &  r n + s m  &= -3 b,~ &  s n &=  -b_0, \nonumber \\ 
    u l &= c^0,~ &  u m + v l &= -3 c,~  &  u n + v m &= -3 d,~ &  v n &= -d_0.
\end{alignat}
Here, we focus on supersymmetric $W=0$ minimum. To stabilize K\"ahler moduli, we will assume non-perturbative dilaton-dependent superpotential $W \sim e^{-aS}$ to realize constant superpotential below the mass scale of axio-dilaton and complex structure moduli. For more details, see, Ref. \cite{Ishiguro:2022pde}.

\medskip

\begin{figure}[H]
\begin{minipage}{0.5\hsize}
  \begin{center}
  \includegraphics[height=100mm]{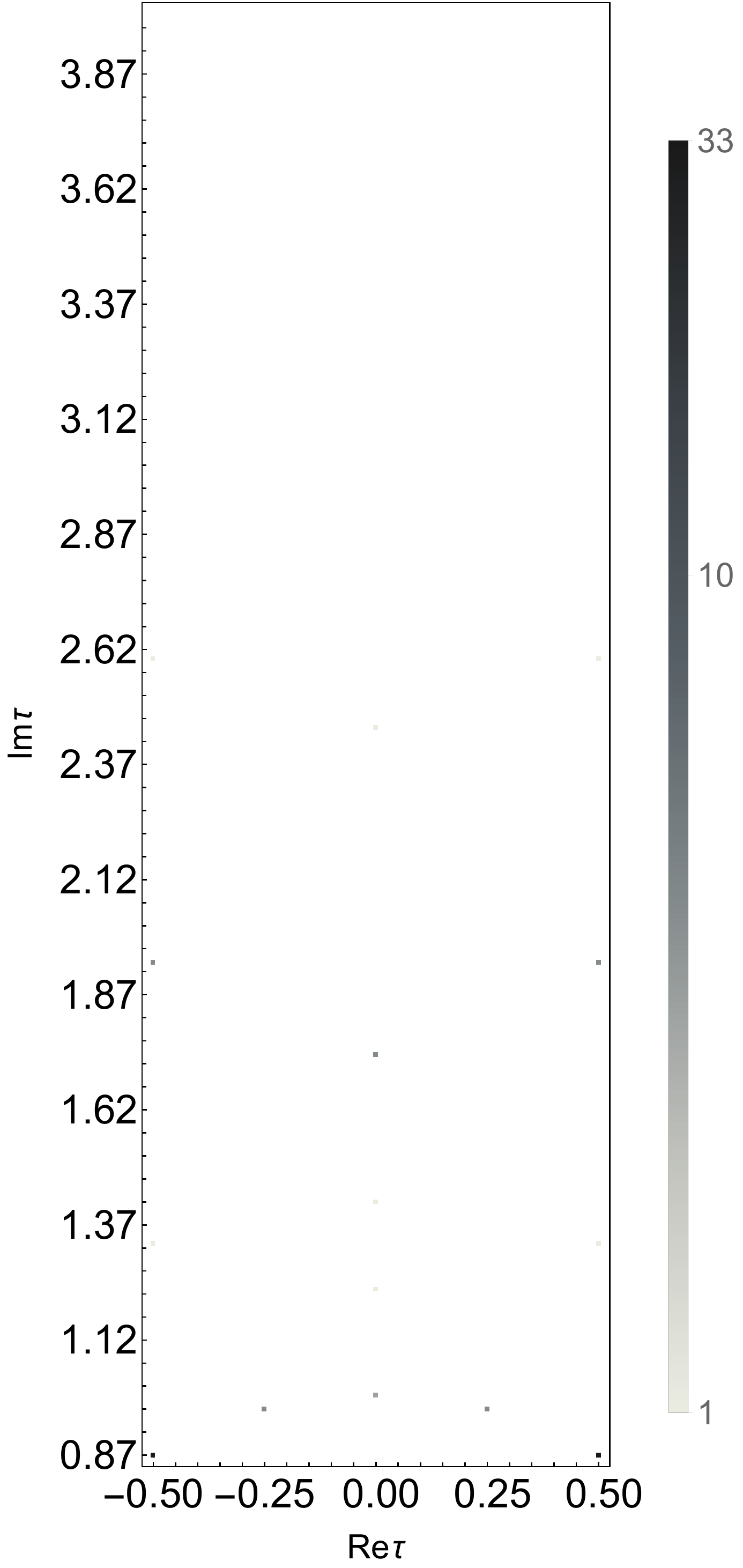}
  \end{center}
 \end{minipage}
 \begin{minipage}{0.5\hsize}
  \begin{center}
   \includegraphics[height=100mm]{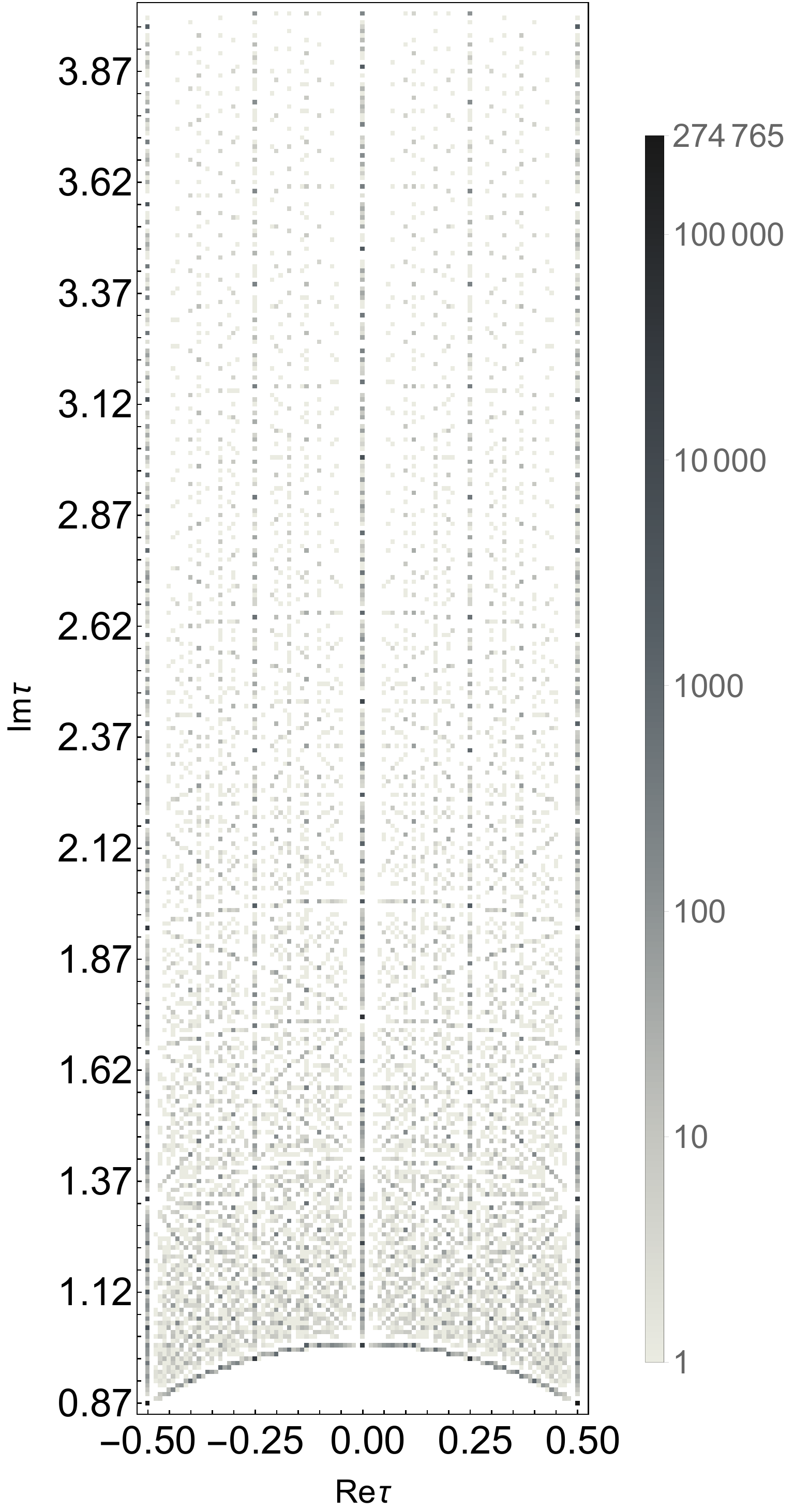}
  \end{center}
 \end{minipage}
  \caption{The numbers of stable flux vacua  on the fundamental domain of $\tau$ 
  for $N_{\rm flux}^{\rm max}=192 \times 10$ in the left panel and  for $N_{\rm flux}^{\rm max}=192 \times 1000$ in the right panel, respectively \cite{Ishiguro:2020tmo}.}
\label{fig:1}
\end{figure}

Since the effective action is invariant under the  $SL(2,\mathbb{Z})_\tau\equiv SL(2,\mathbb{Z})_1= SL(2,\mathbb{Z})_2= SL(2,\mathbb{Z})_3$ and $SL(2,\mathbb{Z})_S$ modular symmetries\footnote{For more details, see, Appendix \ref{app}.}, one can count finite number of physically-distinct flux vacua.\footnote{The modular symmetry was classified in Ref. \cite{Kobayashi:2020hoc} in the context of flux compactifications.} 
Note that we have to be careful about the tadpole cancellation condition of D-brane charges because we deal with a compact manifold. 
In particular, we focus on the cancellation condition of the D3-brane charge, and other conditions will be analyzed in the next subsections. 
Specifically, the flux-induced D3-brane charge
\begin{align}
    N_{\rm flux}&= \frac{1}{l_s^4}\int H_3\wedge F_3 = c^0b_0 -d_0a^0 +\sum_{i=1}^3 (c^ib_i -d_ia^i) = 
    c^0b_0 -d_0a^0 +3(cb -da)\,
    \label{eq:nD3}
\end{align}
satisfies 
\begin{align}
0\leq N_{\rm flux}\leq N_{\rm flux}^{\rm max}={\cal O}(10^5)\,.
    \label{eq:Nflux}
\end{align}
In general, it is difficult to stabilize all the moduli fields including twisted moduli localized at orbifold fixed points in addition to untwisted moduli we focused. 
If Type IIB orientifolds are uplifted to the F-theory in the strong coupling regime, $N_{\rm flux}^{\rm max}={\cal O}(10^5)$ will be a largest value as discussed in Refs.~\cite{Candelas:1997eh,Taylor:2015xtz}. 
In our analysis, we adopt a phenomenological approach such that we simply ignore the concrete tadpole bound and explore the interplay between moduli stabilization and model building. 
This approach allows us to understand the vacuum structure of the string landscape more specifically, as will be shown later. 
Furthermore, each flux quantum is in multiple of 8, that is, $\{a^0, a, b, b_0, c^0, c , d, d_0\}\in 8\, \mathbb{Z}$, and correspondingly $N_{\rm flux} \in 192\,\mathbb{Z}$. 
Since the effective action, as well as the tadpole charge, are invariant under the modular symmetry, 
one can map the moduli VEVs into the fundamental domains. 
The number of stable vacua is shown in Figure \ref{fig:1}, from which there is huge degeneracy at the fixed points in the $SL(2,\mathbb{Z})_\tau$ moduli space. 
In particular, the $\tau = \omega$ vacuum is realized by a high probability such as 62.3 $\%$ for $N_{\rm flux}^{\rm max}=192\times 10$ 
and 40.3 $\%$ for $N_{\rm flux}^{\rm max}=192\times 1000$ \cite{Ishiguro:2020tmo}. 
It can also be justified in a statistical argument. 
By taking the flux quanta as the continuous one, the number of supersymmetric $W=0$ vacua is analytically estimated as \cite{DeWolfe:2004ns}
\begin{align}
    N_{\rm vacua}\sim \frac{\pi^2}{108}\frac{(N_{\rm flux}^{\rm max})^2}{(t(m^2-4ln))^2},
\label{eq:Nvacua}
\end{align}
with 
\begin{align}
    t(x) = \left\{
    \begin{array}{l}
         x\qquad  \,\,\,{\rm for}\,x \equiv 0\quad ({\rm mod}\,3) \\
         3x\qquad {\rm for}\, {\rm otherwise}
    \end{array}
    \right.
    .
\end{align}
Here, ${\rm gcd}(l,m,n)=1$ is adopted in the analysis of Ref. \cite{DeWolfe:2004ns}, but the results are the same with 
our results as pointed out in Ref. \cite{Ishiguro:2020tmo}. 
Remarkably, $\tau = \omega$ corresponding to $(l,m,n)=(1,-1,1)$ is invariant under the discrete $\mathbb{Z}_3$ symmetry, generated by 
\begin{align}
    \{1,\qquad ST,\qquad (ST)^2\},
\end{align}
where $S$ and $T$ are generators of $SL(2,\mathbb{Z})_\tau$:
\begin{align}
    S\,: \tau \rightarrow -\frac{1}{\tau},
    \qquad
    T\,: \tau \rightarrow \tau + 1,
\end{align}
with $(ST)^3=1$. Thus, the effective action in Type IIB flux landscape enjoys the discrete $\mathbb{Z}_3$ symmetry. 
However, it is unclear whether such a $\mathbb{Z}_3$ symmetry still remains in the effective action with the SM spectra. 
In the next section, we will engineer the semi-realistic SM-like models on magnetized D-branes and discuss the role of discrete symmetry.

\subsection{Distribution of $g$-generation models without discrete $B$ field}
\label{sec:generation1}

In addition to O3- and O7-planes located at fixed loci, 
we construct semi-realistic models on $N_a$ stacks of magnetized D$(3+2n)$-branes wrapping $2n$-cycles on 
$T^6/(\mathbb{Z}_2\times \mathbb{Z}_2^\prime)$ orientifolds. 
We turn on the background $U(1)_a$ gauge field strength $F_a$ on $(T^2)_i$, 
\begin{align}
    \frac{m_a^i}{2\pi} \int_{T_i^2} F_a^i = n_a^i, 
\end{align}
where wrapping numbers of $N_a$ D$(3+2n)$-branes on $(T^2)_i$ are represented by integers $m_a^i$ 
with non-vanishing 0, 1, 2 and, 3 values on D3-, D5-, D7-, D9-branes, respectively. 
Note that $\{n_a^i,m_a^i\}$ for each $a$ and $i$ are assumed to be coprime numbers, and 
only the wrapping number $m_a^i$ transforms as 
$\Omega {\cal R}\,:\,m_a^i\rightarrow -m_a^i$ under $\Omega{\cal R}$. 
For practical purposes, let us introduce the homology classes of each $(T^2)_i$, that is, 
$[{\bf 0}]_i$ and $[{\bf T^2}]_i$ for the class of point and of the two-torus with 
$[{\bf 0}]_i \cdot [{\bf T^2}]_i = - [{\bf T^2}]_i \cdot [{\bf 0}]_i = 1$. 
Then, the stack $a$ of D-branes has an associated homology class:
\begin{align}
    [{\bf Q}_a] = \Pi_{i=1}^3 \left(n_a^i [{\bf 0}]_i + m_a^i [{\bf T^2}]_i  \right).
\end{align}
Similarly, the 64 O3- and 4 O$7_i$ planes are expressed by RR charges -32 times the 
following homology classes:
\begin{align}
    [{\bf Q}_{O3}] &= [{\bf 0}]_1 \times [{\bf 0}]_2 \times [{\bf 0}]_3,\qquad
    [{\bf Q}_{O7_1}] = -[{\bf 0}]_1 \times [{\bf T^2}]_2 \times [{\bf T^2}]_3,
    \nonumber\\
    [{\bf Q}_{O7_2}] &= -[{\bf T^2}]_1 \times [{\bf 0}]_2 \times [{\bf T^2}]_3,\qquad
    [{\bf Q}_{O7_3}] = -[{\bf T^2}]_1 \times [{\bf T^2}]_2 \times [{\bf 0}]_3.
\end{align}

\medskip

Remarkably, these gauge fluxes will lead to semi-realistic D-brane models, 
that is, gauge groups $G_{\rm SM} \times G'$ with chiral spectra. 
In particular, the index theorem tells us that the number of chiral zero-modes 
between two stacks $a$ and $b$ of D-branes on $T^6 = (T^2)_1\times (T^2)_2\times (T^2)_3$ is counted by
\begin{align}
    I_{ab} = [{\bf Q}_a]\cdot [{\bf Q}_b] = \Pi_{i=1}^3 (n_a^i m_b^i -n_b^i m_a^i).
    \label{eq:Index}
\end{align}
However, some of the couplings of zero-modes are projected out by $\mathbb{Z}_2\times \mathbb{Z}_2'$ projection (\ref{eq:projections}). 
Indeed, internal fermionic wavefunctions transform as
\begin{align}
    \theta\,&:\,\psi(z_1,z_2,z_3) \rightarrow s_1s_2 \psi(-z_1,-z_2,z_3),
    \nonumber\\
    \theta'\,&:\,\psi(z_1,z_2,z_3) \rightarrow s_2s_3 \psi(z_1,-z_2,-z_3),
\end{align}
where $s_i = {\rm sign}(I_{ab}^i)$ corresponds to the chirality on each torus and 
its product ($s_1s_2s_3$) corresponds to the 4D chirality. 
Thus, there exist $\mathbb{Z}_2$-even and -odd modes on each torus whose explicit 
form of zero-mode wavefunctions is shown later. 
Note that the two conditions should be consistent with each other; that is,
\begin{align}
    \psi(z_1,z_2,z_3) = s_1s_2 \psi(-z_1,-z_2,z_3) = s_2s_3 \psi(z_1,-z_2,-z_3),
\end{align}
from which allowed zero-modes are given by a specific combination of $\mathbb{Z}_2$-even modes ($\psi_{\rm even}^i$) and $\mathbb{Z}_2$-odd zero-modes ($\psi_{\rm odd}^i$) on $(T^2)_i$:
\begin{align}
    \psi = \{ &\psi_{\rm even}^i\psi_{\rm even}^j\psi_{\rm even}^k,\,\,
    \psi_{\rm even}^i\psi_{\rm even}^j\psi_{\rm odd}^k,\,\,
    \psi_{\rm even}^i\psi_{\rm odd}^j\psi_{\rm even}^k,\,\,
    \psi_{\rm odd}^i\psi_{\rm even}^j\psi_{\rm even}^k,
    \nonumber\\
    &\psi_{\rm odd}^i\psi_{\rm odd}^j\psi_{\rm odd}^k,\,\,
    \psi_{\rm odd}^i\psi_{\rm odd}^j\psi_{\rm even}^k,\,\,
    \psi_{\rm odd}^i\psi_{\rm even}^j\psi_{\rm odd}^k,\,\,
    \psi_{\rm even}^i\psi_{\rm odd}^j\psi_{\rm odd}^k\},
\end{align}
with $i\neq j \neq k$. 
Since the number of these $\mathbb{Z}_2$-even and -odd zero-modes is counted by \cite{Marchesano:2013ega}
\begin{align}
    I_{\rm even}^i = \frac{1}{2}(I_{ab}^i +s_i f_i), 
    \qquad
    I_{\rm odd}^i = \frac{1}{2}(I_{ab}^i - s_i f_i),    
    \label{eq:index_evenodd}
\end{align}
with $f_i = 1$ for odd $I_{ab}^i$ and $f_i = 2$ for even $I_{ab}^i$, the total number of zero-modes is still described by 
\begin{align}
    I_{ab}= \Pi_{i=1}^3 (I_{\rm even}^i + I_{\rm odd}^i).
\end{align}
Here, we assume $I_{ab}^i\neq 0$. 
If one of the indices is 0, e.g., $I_{ab}^3=0$, the spectrum is not chiral, and the index is counted by \cite{Marchesano:2013ega}
\begin{align}
    I_{ab} = 
    \left\{
    \begin{array}{c}
I_{\rm even}^1I_{\rm even}^2 - I_{\rm odd}^1I_{\rm odd}^2\qquad (s_1>0,\,s_2>0)\\
I_{\rm even}^1I_{\rm odd}^2 - I_{\rm odd}^1I_{\rm even}^2\qquad (s_1>0,\,s_2<0)\\
I_{\rm odd}^1I_{\rm even}^2 - I_{\rm even}^1I_{\rm odd}^2\qquad (s_1<0,\,s_2>0)\\
I_{\rm odd}^1 I_{\rm odd}^2 - I_{\rm even}^1 I_{\rm even}^2\qquad (s_1<0,\,s_2<0)\\
    \end{array}
    \right.
    \label{eq:index_Higgs}
    .
\end{align}

\medskip

On $N_a$ stack of D-branes that does not lie on one of the O-planes, 
the mass spectra consist of $U(N_a/2)$ vector multiplets and three adjoint chiral multiplets (called a $aa$ sector)\footnote{The 
$U(N_a)$ gauge group reduces to $U(N_a/2)$ due to the orbifold projection.}. 
On the other hand, when $2N_a$ stack of D-branes lies on one of the O-planes, 
the mass spectra consist of $USp(N_a)$ vector multiplets and three antisymmetric chiral multiplets, which we also call a $aa$ sector. 
In addition, there are chiral multiplets that arise from intersections of two different stacks $a$ and $b$ of D-branes or the stack $a$ and its orientifold image 
$a'$, as summarized in Table \ref{tab:spectra}. 
\begin{table}[H]
\centering
\begin{tabular}{|c|c|c|}\hline
Sectors & Representations                       & Multiplicities                     \\\hline
$ab+ ba$ & $(\fund_a, \antifund_b) $ & $I_{ab}$                          \\
$ab'+ b'a$ & $(\fund_a, \fund_b)$       & $I_{ab'}$                         \\
$aa'+ a'a$ &  $\Ysymm$ (symmetric)   & $\frac{1}{2}(I_{aa'} - 4I_{aO})$ \\
$aa'+ a'a$ & $\Yasymm$ (anti-symmetric)     & $\frac{1}{2}(I_{aa'} + 4I_{aO})$ \\\hline
\end{tabular}
\caption{Multiplicities of chiral zero-modes in each sector.}
\label{tab:spectra}
\end{table}

\medskip

Since the magnetic fluxes induce the D3- and D7-brane charges, 
we have to be careful about their tadpole cancellation conditions:
\begin{align}
    D3\;&:\; \sum_a N_a n_a^1n_a^2n_a^3 +\frac{1}{2}N_{\rm flux} =16,
    \nonumber\\
    D7_1\;&:\; \sum_a N_a n_a^1m_a^2m_a^3 =-16,
    \nonumber\\
    D7_2\;&:\; \sum_a N_a n_a^2m_a^1m_a^3 =-16,
    \nonumber\\
    D7_3\;&:\; \sum_a N_a n_a^3m_a^1m_a^2 =-16.
    \label{eq:Tad-Dbranes}
\end{align}
If there exists D9-branes with constant magnetic fluxes, they are mapped to anti D9-branes with the opposite magnetic fluxes under the orientifold involution. Thus, D9-brane tadpole charges are canceled. Similar things happen for D5-branes with constant magnetic fluxes as well.
These conditions play a role of the cancellation of 4D chiral anomalies, 
but K-theory conditions require extra constraints. 
Indeed, probe D3 and D7-branes with $USp(2) \simeq SU(2)$ gauge group suffer from a global gauge anomaly if the number of 4D fermions charged in the fundamental representation of $SU(2)$ is odd \cite{Witten:1982fp}. 
It imposes the following K-theory constraints \cite{Uranga:2000xp}:
\begin{align}
    \left\{\sum_a N_a n_a^1n_a^2n_a^3,\,\sum_a N_a n_a^1m_a^2m_a^3,\, 
    \sum_a N_a n_a^2m_a^1m_a^3, \, \sum_a N_a n_a^3m_a^1m_a^2\right\}\in 4\,\mathbb{Z}.
    \label{eq:Ktheory}
\end{align}

\medskip

Since magnetized D9-branes with negative $n_{1,2,3}$ will carry anti D3- and D7-brane charges, 
it will be possible to construct semi-realistic $3$-generation models on the flux background (see, e.g., \cite{Marchesano:2004xz,Kumar:2005hf}). 
In these analyses, we have not introduced anti-D3 branes satisfying tadpole cancellation condition, but it would be 
possible to construct realistic models, taking into account the effect of anti-D3 brane annihilations with flux \cite{Kachru:2002gs}. 
Note that the ${\cal N}=1$ supersymmetry on the orientifold background will be preserved when the following 
condition is satisfied \cite{Cvetic:2005bn}:
\begin{align}
    \sum_i \biggl[ \tan^{-1} \left( \frac{m_a^i}{n_a^i}{\cal A}_i\right) + \theta(n_a^i)\pi \biggl] = 0
    \qquad ({\rm mod}\,2\pi),
    \label{eq:SUSY}
\end{align}
with
\begin{align}
    \theta (n_a^i) =
    \left\{
    \begin{array}{l}
         0\quad (n_a^i \geq 0)  \\
         1\quad (n_a^i <0) 
    \end{array}
    \right.
    ,
\end{align}
where ${\cal A}_i$ denote the area of the torus $(T^2)_i$. 

\medskip

\begin{figure}[H]
\begin{minipage}{0.5\hsize}
  \begin{center}
   \includegraphics[height=44mm]{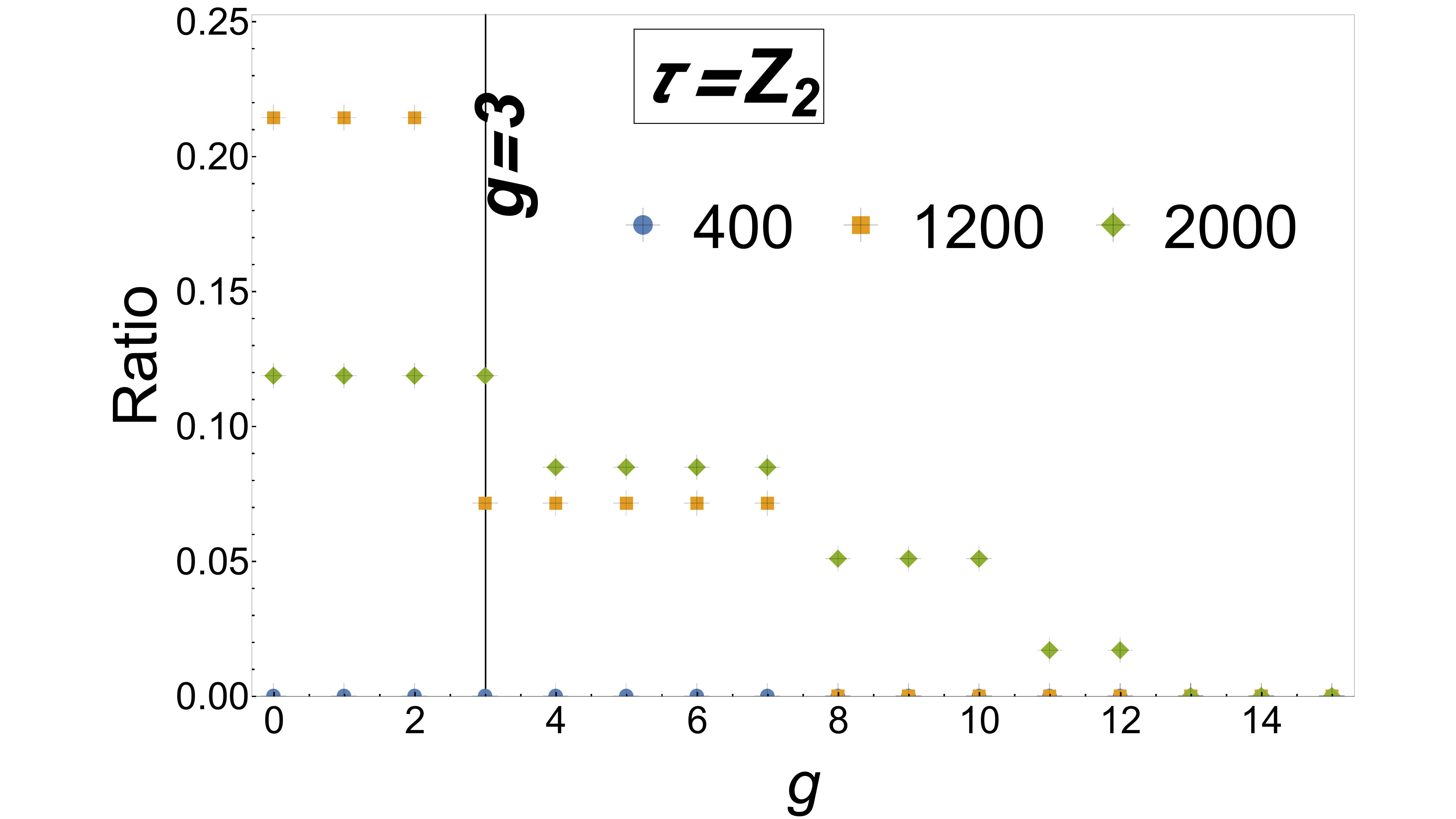}
  \end{center}
 \end{minipage}
 \begin{minipage}{0.5\hsize}
  \begin{center}
   \includegraphics[height=44mm]{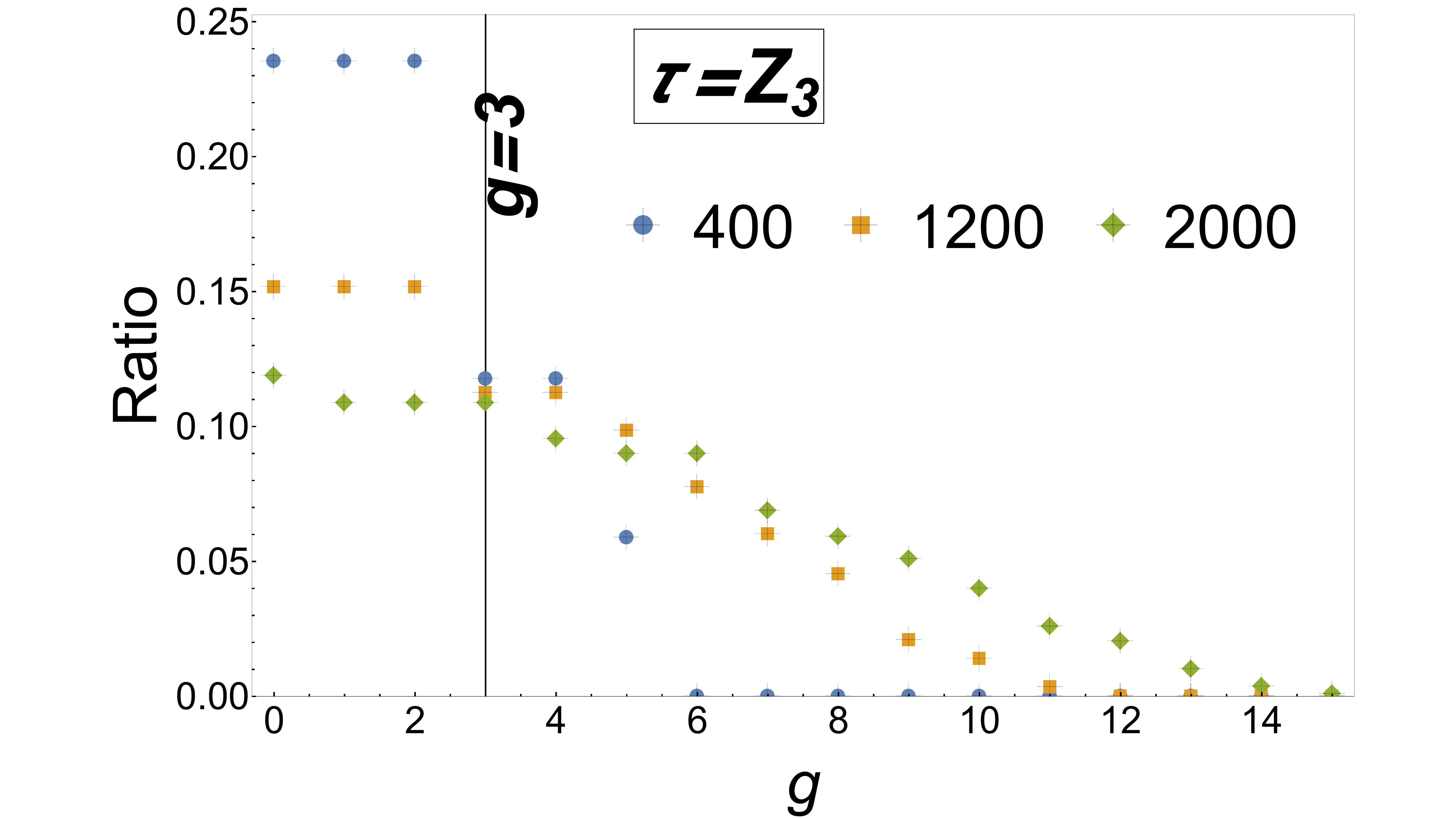}
  \end{center}
 \end{minipage}
  \caption{The numbers of models as a function of the generation number $g$ at $\tau = i$ and 
$\tau=\omega$, respectively. Note that there exists $\mathbb{Z}_2$ symmetry at $\tau =i$ generated by $\{1,S\}$. Here, the vertical axis represents the ratio of the number of models to the total number of models. There are three plots in each panel, and each of them corresponds to the maximum value of the D3-brane charge 
$|Q^{\rm hid}_{D3}| = 400, 1200, 2000$.}
\label{fig:model1_g}
\end{figure}

\begin{figure}[H]
\begin{minipage}{0.5\hsize}
  \begin{center}
   \includegraphics[height=110mm]{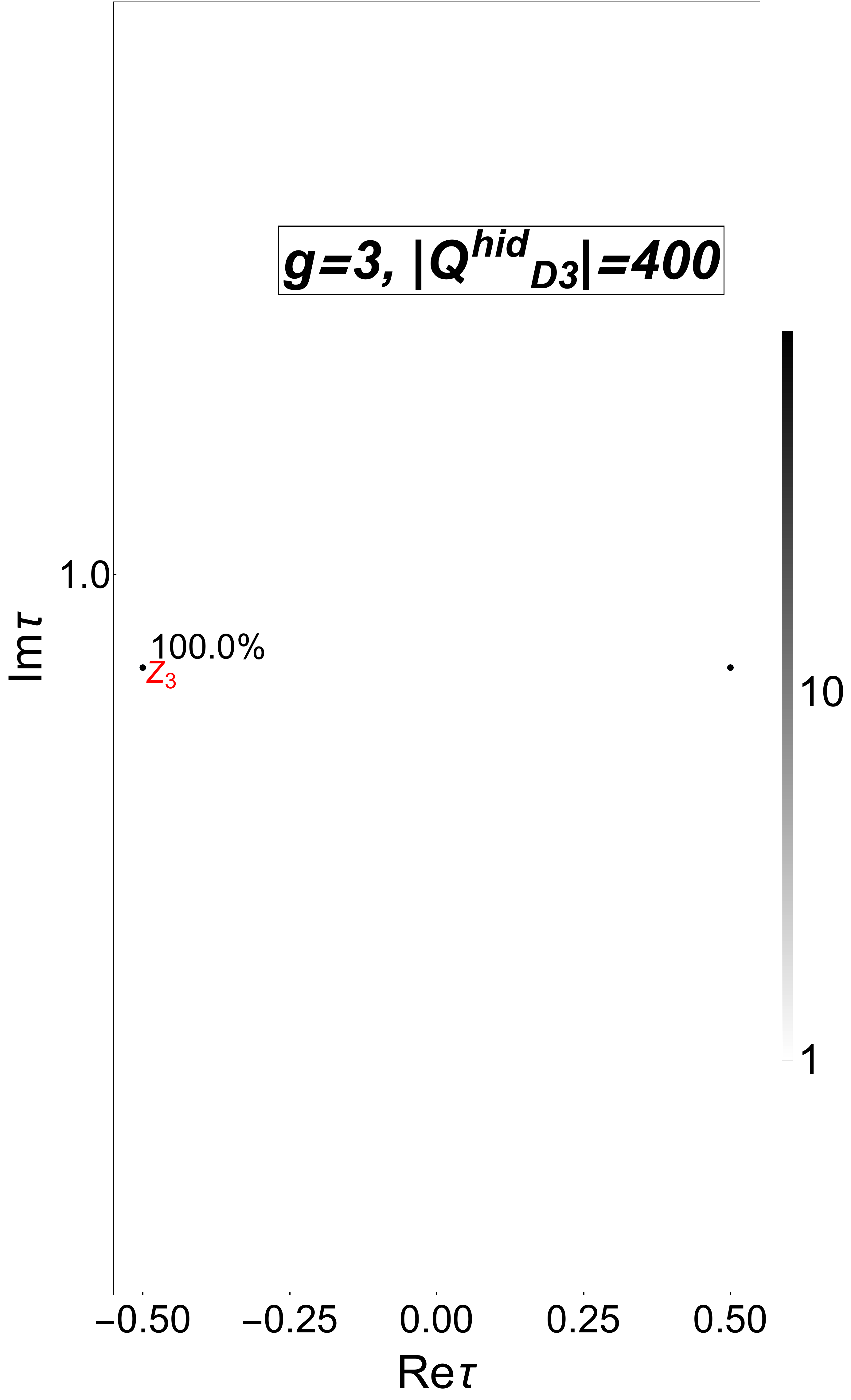}
  \end{center}
 \end{minipage}
 \begin{minipage}{0.5\hsize}
  \begin{center}

   \includegraphics[height=110mm]{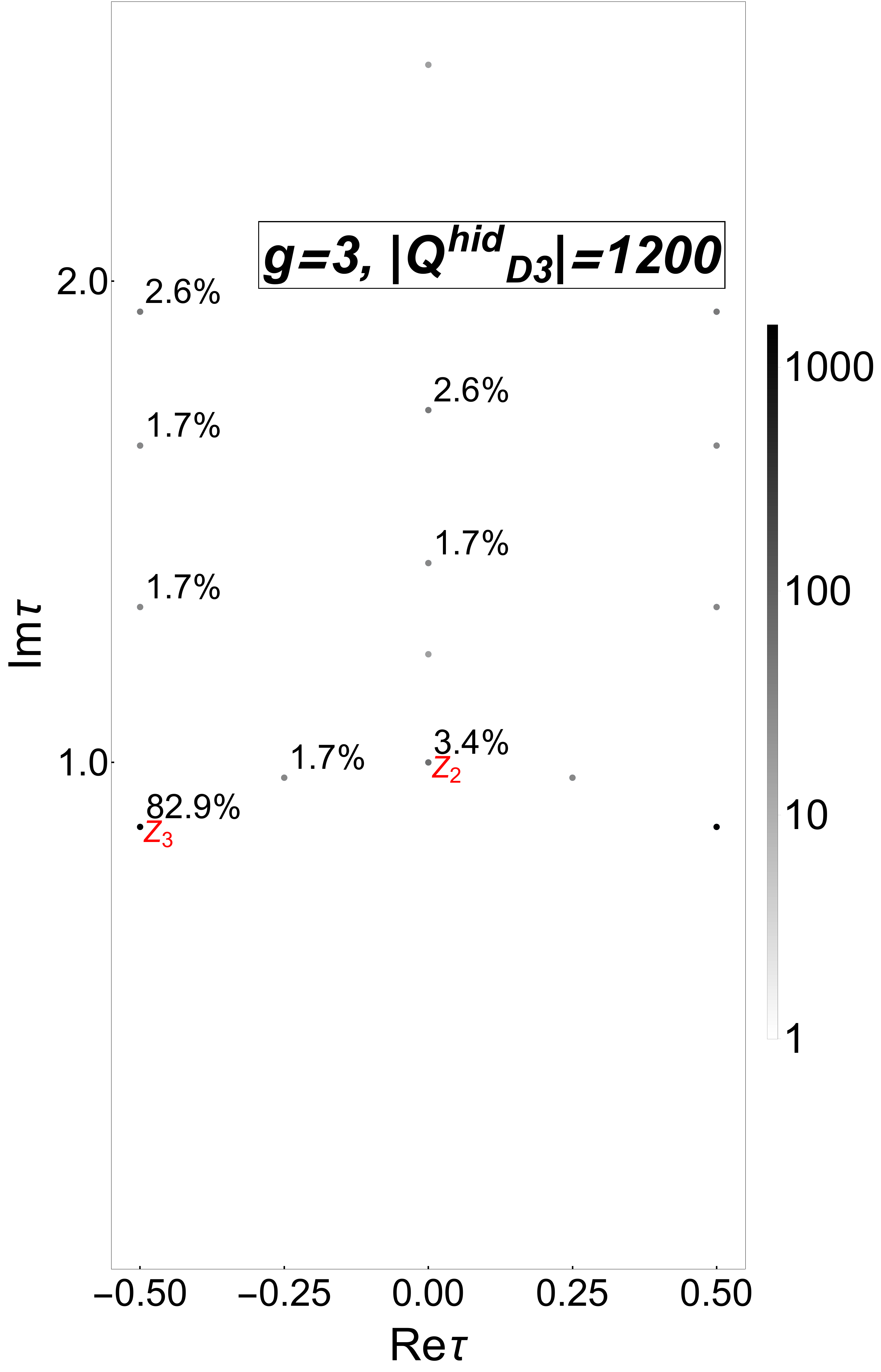}
  \end{center}
 \end{minipage}
  \caption{The numbers of stable flux vacua with $g=3$ generation of quarks/leptons on the fundamental domain of $\tau$ for the maximum value of D3-brane charge $Q_{D3}^{\rm hid}\bigl|_{\rm max}=400$ in the left panel and for $Q_{D3}^{\rm hid}\bigl|_{\rm max}=1200$ in the right panel, respectively.}
\label{fig:model1_fund}
\end{figure}

For concreteness, let us consider the local brane configurations with SM spectra as shown in Table \ref{tab:model1} \cite{Marchesano:2004xz}, leading to $g$ generation of quarks and leptons $I_{ab}=I_{ac}=g$.\footnote{Here and in what follows, we call the generation number the sum of $\mathbb{Z}_2$-even and -odd modes without specifying it.} 
\begin{table}[H]
    \centering
    \begin{tabular}{|c|c|c|c|c|} \hline
        $N_\alpha$ & {\rm Gauge~group} & ($n_\alpha^1, m_\alpha^1$) & ($n_\alpha^2, m_\alpha^2$) & ($n_\alpha^3, m_\alpha^3$)\\ \hline \hline
        $N_a=6$ & $SU(3)_C$ & (1,0) & ($g, 1$) & ($g, -1$)\\ \hline
         $N_b=2$ & $USp(2)_L$ & (0,1) & ($1, 0$) & ($0, -1$)\\ \hline
        $N_c=2$ & $USp(2)_R$ & (0,1) & ($0, -1$) & ($1,0$)\\ \hline
        $N_d=2$ & $U(1)_d$ & (1,0) & ($g, 1$) & ($g, -1$)\\ \hline
    \end{tabular}
    \caption{D-brane configurations leading to left-right symmetric Minimal Supersymmetric Standard Model (MSSM). 
    The magnetic flux $g$ determines the generations of quark and lepton chiral multiplets in the visible sector.}
    \label{tab:model1}
\end{table}
The supersymmetry condition (\ref{eq:SUSY}) is satisfied when
\begin{align}
    {\cal A}_2 = {\cal A}_3.
\end{align}
Furthermore, some of $U(1)s$ become massive by absorbing axions associated with Ramond-Ramond fields through the Green-Schwarz mechanism. 
Indeed, the dimensional reduction of the Chern-Simons couplings in the D-brane action 
induces the corresponding 4D couplings:
\begin{align}
   &N_a n_a^1n_a^2 n_a^3 \int_{R^{1,3}} C_2 \wedge F_a,
   \qquad
   N_a m_a^i n_a^j n_a^k \int_{R^{1,3}} C_2^i \wedge F_a,
   \nonumber\\
   &N_a m_a^j m_a^k n_a^i \int_{R^{1,3}} {\cal C}_2^i \wedge F_a,
   \qquad
   N_a m_a^1 m_a^2 m_a^3 \int_{R^{1,3}} {\cal C}_2 \wedge F_a,
\label{eq:GS}
\end{align}
with
\begin{align}
    C_2^i := \int_{(T^2)_i} C_4,\qquad
    {\cal C}_2^i := \int_{(T^2)_j \times (T^2)_k} C_6\qquad
    {\cal C}_2 := \int_{(T^6)} C_8.
\end{align}
To satisfy the tadpole cancellation conditions, we have also supposed the existence of magnetized D9-branes to satisfy the tadpole cancellation conditions. 
It means that the D3-brane charge induced by the magnetic flux on D9-branes ${\cal Q}_{D3}^{\rm hid}$ satisfy
\begin{align}
    {\cal Q}_{D3}^{\rm hid}  =16- \frac{N_{\rm flux}}{2}-8g^2, 
    \leftrightarrow
    8g^2 =  -{\cal Q}_{D3}^{\rm hid}  +16- \frac{N_{\rm flux}}{2}.
    \label{eq:Tad-Fluxes}
\end{align}
Since there are several possibilities for the choice of magnetized D9-brane sectors,
we freely change the value of ${\cal Q}_{D3}^{\rm hid}$ to reveal the mutual relation between the generation number $g$ and the flux quanta $N_{\rm flux}$.\footnote{See, 
e.g., \cite{Kumar:2005hf}, for the numerical analysis searching ${\cal Q}_{D3}^{\rm hid}$.} 
In Fig. \ref{fig:model1_g}, we change the maximum value of ${\cal Q}_{D3}^{\rm hid}$ as $|{\cal Q}_{D3}^{\rm hid}|= 400, 1200, 2000$, each which we analyze the distribution of flux vacua at fixed $\tau$ with respect to $g$. 
It turns out that the number of flux vacua increases when $g$ is smaller. 
Thus, the small generation number is favored in the string landscape. 
Furthermore, when we restrict ourselves to three-generation models, that is, $g=3$, 
left-right MSSM-like models are still peaked at the $\mathbb{Z}_3$ fixed point $\tau = \omega$ as shown in Fig. \ref{fig:model1_fund}. 
This phenomenon is similar to the analysis of Sec. \ref{sec:moduli}, 
but the percentage of three-generation clustered regions differs from before. 
One can further study the Yukawa couplings derived in Type IIB magnetized D-brane models \cite{Cremades:2004wa}. 
In the current brane configuration, the Yukawa couplings of quarks and leptons are rank one, and the flavor structure is trivial due to the fact that the flavor structure is realized from two different tori. 
Thus, we move on to the other magnetized D-brane model, inducing the non-trivial flavor structure of quarks and leptons.

\subsection{Distribution of $g$-generation models with discrete $B$ field}
\label{sec:generation2}

In this section, we add a discrete value of Kalb-Ramond $B$-field along one of the two-tori \cite{Blumenhagen:2000ea}, in particular, $(T^2)_3$, corresponding to the twisted torus in the T-dual IIA string theory. 
Since $B$-field induces the half-integer flux, the magnetic flux on the third torus is modified as $\tilde{n}_a^3 = n_a^3 + \frac{1}{2}m_a^3$. 
According to it, the tadpole cancellation conditions are given by \cite{Cascales:2003zp}\footnote{
When we consider a different tilted direction, the effective flux is given by $\tilde{m}_a^3 = m_a^3 + \frac{1}{2}n_a^3$ as discussed in the T-dual IIA side \cite{Cvetic:2001nr}.}
\begin{align}
    D3\;&:\; \sum_a N_a n_a^1n_a^2\tilde{n}_a^3 +\frac{1}{2}N_{\rm flux} = 8,
    \nonumber\\
    D7_1\;&:\; \sum_a N_a n_a^1m_a^2 m_a^3 =-16,
    \nonumber\\
    D7_2\;&:\; \sum_a N_a n_a^2m_a^1 m_a^3 =-16,
    \nonumber\\
    D7_3\;&:\; \sum_a N_a \tilde{n}_a^3m_a^1m_a^2 =-8.
    \label{eq:Tad_tilt}
\end{align}
The cancellations of D5- and D9-brane charges are realized as mentioned below Eq. (\ref{eq:Tad-Dbranes}).
The other SUSY condition (\ref{eq:SUSY}) and K-theory condition (\ref{eq:Ktheory}) are also written in terms of $\tilde{n}_a^3$. 
For concreteness, let us consider the local brane configurations with SM spectra as shown in Table \ref{tab:model2}, leading to $g$ generation of quarks and leptons $I_{ab}=I_{ca}=g$.\footnote{Similar brane configurations are discussed in T-dual Type IIA string theory, e.g., \cite{Chen:2007px}.} The supersymmetry condition (\ref{eq:SUSY}) is satisfied when
\begin{align}
    g{\cal A}_1 = {\cal A}_2 = 2{\cal A}_3.
\end{align}
In this model, there $U(1)$s in the gauge symmetry $U(4)_C\times U(2)_L\times U(2)_R$ absorb axions through the Green-Schwarz couplings (\ref{eq:GS}). The remaining gauge symmetry is described by $SU(4)_C\times SU(2)_L\times SU(2)_R$. 
Furthermore, the Pati-Salam gauge symmetry can be broken to MSSM gauge group by the splitting of $a$ and $c$ stack of D-branes, but we leave the detailed study of open string moduli for future work. 
\begin{table}[H]
    \centering
    \begin{tabular}{|c|c|c|c|c|} \hline
        $N_\alpha$ & {\rm Gauge~group} & ($n_\alpha^1, m_\alpha^1$) & ($n_\alpha^2, m_\alpha^2$) & ($\tilde{n}_\alpha^3, m_\alpha^3$)\\ \hline \hline
        $N_a=8$ & $U(4)_C$ & ($0,-1$) & ($1, 1$) & ($1/2, 1$)\\ \hline
         $N_b=4$ & $U(2)_L$ & ($g,1$) & ($1, 0$) & ($1/2, -1$)\\ \hline
        $N_c=4$ & $U(2)_R$ & ($g,-1$) & ($0, 1$) & ($1/2, -1$)\\ \hline
    \end{tabular}
    \caption{D-brane configurations leading to Pati-Salam-like model. 
    The magnetic flux $g$ determines the generations of quark and chiral chiral multiplets in the visible sector, where $\tilde{n}=n+m/2$.}
    \label{tab:model2}
\end{table}

\medskip

\begin{figure}[H]
\begin{minipage}{0.5\hsize}
  \begin{center}
   \includegraphics[height=44mm]{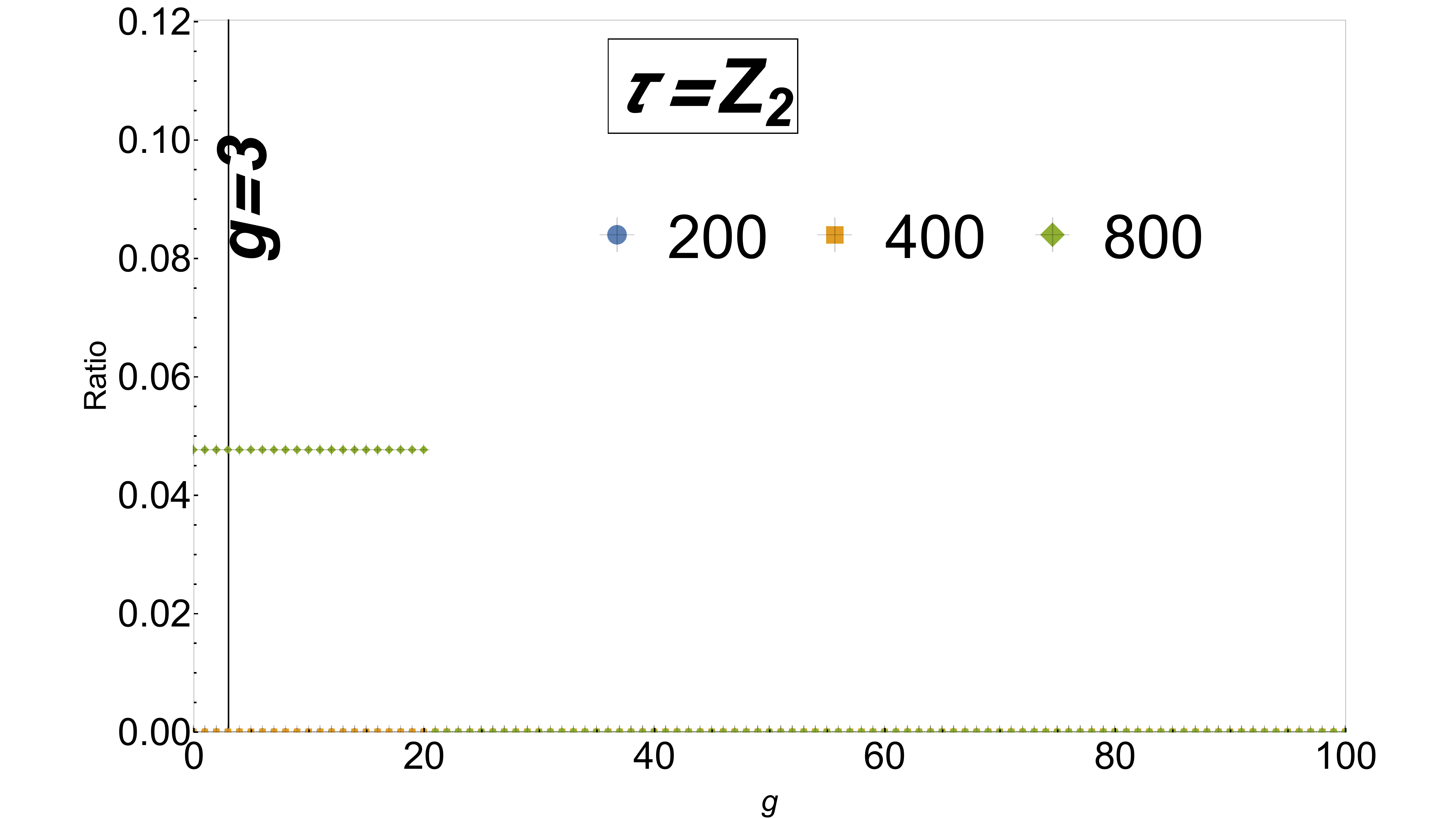}
  \end{center}
 \end{minipage}
 \begin{minipage}{0.5\hsize}
  \begin{center}

   \includegraphics[height=44mm]{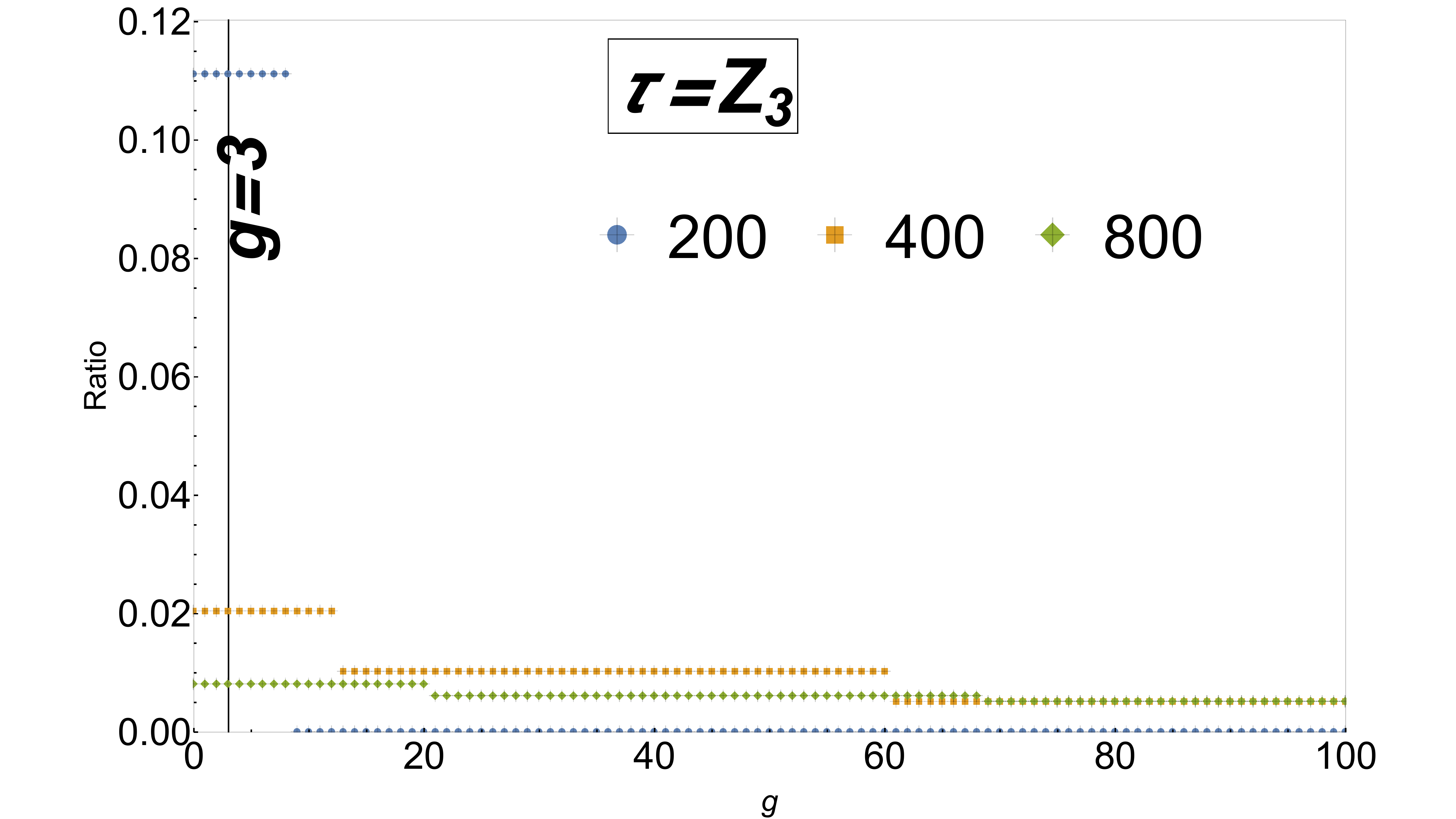}
  \end{center}
 \end{minipage}
  \caption{The numbers of models as a function of the generation number $g$ at $\tau = i$ and 
$\tau=\omega$, respectively. Here, the vertical axis represents the ratio of the number of models to the total number of models. There are three plots in each panel, and each of them corresponds to the maximum value of the D3-brane charge $|Q^{\rm hid}_{D3}| = 200, 400, 800$.}
\label{fig:model2_g}
\end{figure}

For the same reason as the analysis of the previous section, we allow several values of ${\cal Q}_{D3}^{\rm hid}$ satisfying 
\begin{align}
    {\cal Q}_{D3}^{\rm hid}  =8- \frac{N_{\rm flux}}{2} - 2g, 
    \leftrightarrow
    2g =  -{\cal Q}_{D3}^{\rm hid}  + 8- \frac{N_{\rm flux}}{2},
    \label{eq:Tad2}
\end{align}
to reveal the mutual relation between the generation number $g$ and the flux quanta $N_{\rm flux}$. 
In Fig. \ref{fig:model2_g}, we change the maximum value of ${\cal Q}_{D3}^{\rm hid}$ as $|{\cal Q}_{D3}^{\rm hid}|=200, 400, 800$, each which we analyze the distribution of flux vacua at fixed $\tau$ with respect to $g$. 
It turns out that the number of flux vacua also increases when $g$ is smaller, although the behavior is different from 
the previous analysis. 
Thus, the string landscape leads to the small generation number. 
Furthermore, when we restrict ourselves to three-generation model, that is, $g=3$, 
Pati-Salam models are still peaked at the $\mathbb{Z}_3$ fixed point $\tau = \omega$ in a similar to 
the analysis of Sec. \ref{sec:moduli}. 
In contrast to the previous models in Sec. \ref{sec:generation1}, 
the Yukawa couplings of quarks and leptons are rank 3 and the flavor structure is non-trivial due to the fact that the flavor structure is originated from one of tori. 
We will discuss the relation between flavor symmetries and modular symmetries in the next section.

\begin{figure}[H]
\begin{minipage}{0.5\hsize}
  \begin{center}
   \includegraphics[height=110mm]{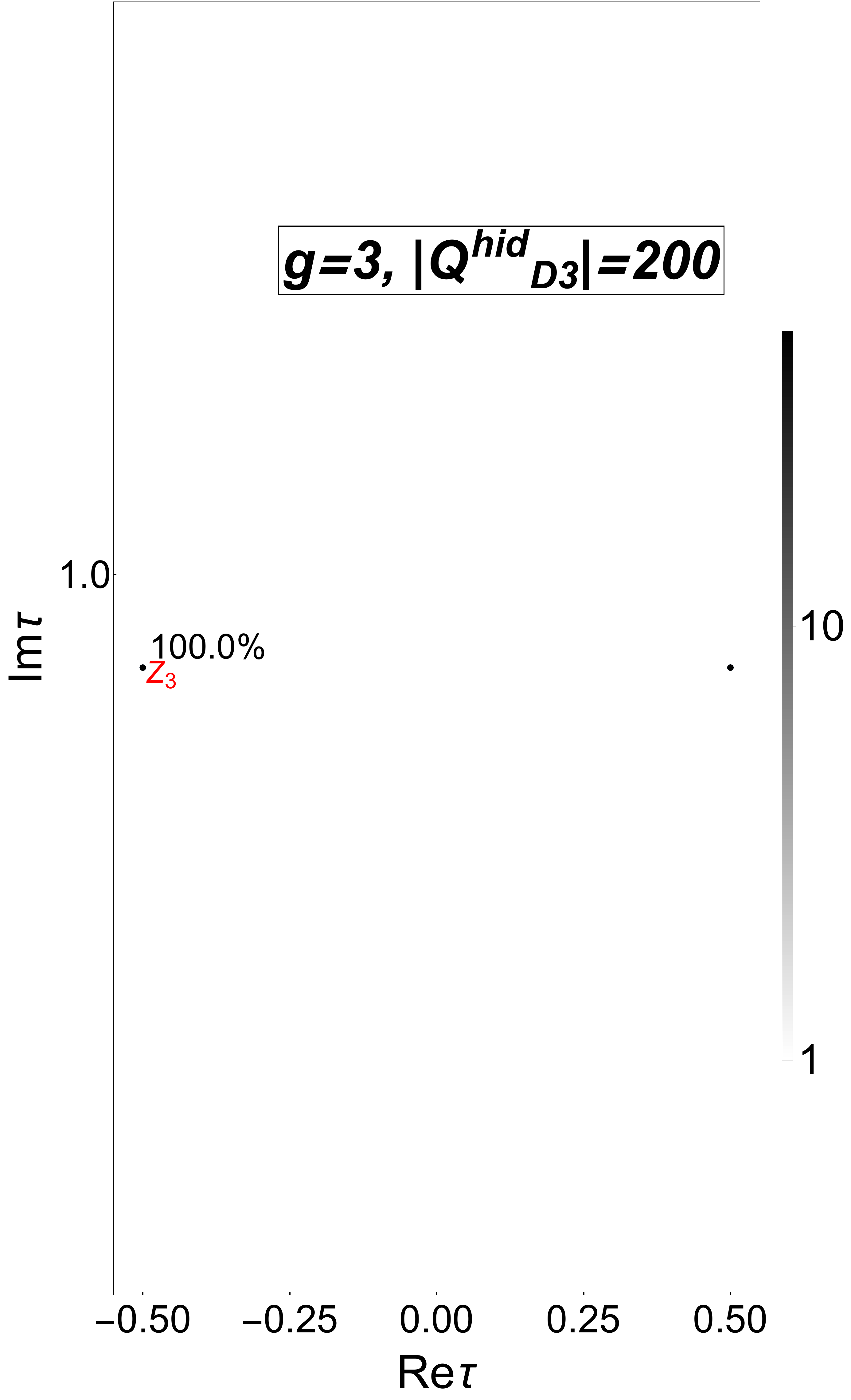}
  \end{center}
 \end{minipage}
 \begin{minipage}{0.5\hsize}
  \begin{center}

   \includegraphics[height=110mm]{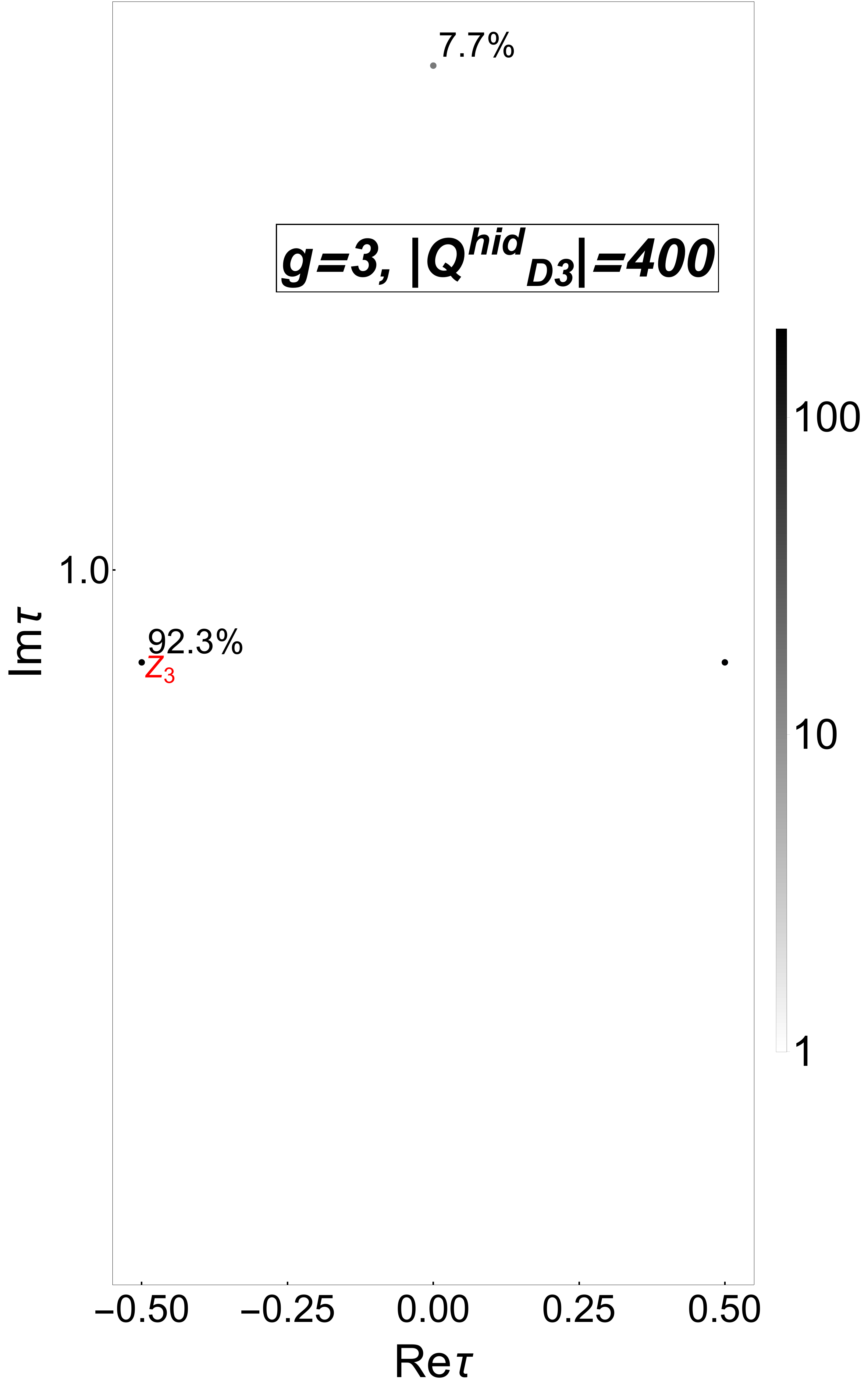}
  \end{center}
 \end{minipage}
  \caption{The numbers of stable flux vacua with $g=3$ generation of quarks/leptons on the fundamental domain of $\tau$ for the maximum value of D3-brane charge $Q_{D3}^{\rm hid}\bigl|_{\rm max}=200$ in the left panel and for $Q_{D3}^{\rm hid}\bigl|_{\rm max}=400$ in the right panel, respectively.}
\label{fig:model2_fund}
\end{figure}

\section{Eclectic Flavor Symmetry in Type IIB flux vacua}
\label{sec:eclectic}

So far, we have studied distribution of moduli fields and remaining modular symmetry in the low-energy effective action. 
In this section, we discuss a flavor and CP symmetries of degenerate chiral zero-modes on D-branes and its relation to modular symmetry. 

In Secs. \ref{sec:T2} and \ref{sec:T2Z2}, we show that the metaplectic modular symmetry introduced in Sec. \ref{sec:meta} is useful to describe the matter wavefunctions and Yukawa couplings on $T^2$ and $T^2/\mathbb{Z}_2$ with magnetic fluxes in an uniform way. 
Remarkably, the CP symmetry can be regarded as an outer automorphism of the modular symmetry as discussed in Sec. \ref{sec:CP}. 
These 6D bottom-up models can be embedded in 10D Type IIB magnetized D-brane models with stabilized moduli. In Sec. \ref{sec:10D}, we discuss the metaplectic modular flavor symmetries together with 
traditional flavor and CP symmetries in the framework of eclectic symmetry.

\subsection{Metaplectic modular symmetry}
\label{sec:meta}

Since the Yukawa couplings of quarks and leptons are described by a half-integer 
modular form, they are formulated in the context of metaplectic group $Mp(2, \mathbb{Z})$. 
Following Ref. \cite{Liu:2020msy}, let us briefly review the notion of $Mp(2, \mathbb{Z})$ which is a twofold covering group of $SL(2, \mathbb{Z})$.
\begin{align}
    Mp(2,\mathbb{Z}) = \left\{
    \tilde{\gamma} = (\gamma, \varphi(\gamma, \tau))\, \Biggl|
    \,\,\gamma = 
    \left(
    \begin{array}{cc}
        a & b \\
        c & d
    \end{array}
    \right)
    \in SL(2,\mathbb{Z}),
    \quad
    \varphi(\gamma, \tau)^2 = (c\tau +d)
    \right\}
    ,
\end{align}
where the multiplication law is defined as
\begin{align}
    (\gamma_1, \varphi(\gamma_1, \tau))(\gamma_2, \varphi(\gamma_2, \tau))
    = (\gamma_1 \gamma_2, \varphi(\gamma_1, \gamma_2\tau)\varphi(\gamma_2,\tau)).
\end{align}
When we redefine $\varphi(\gamma, \tau) = \pm (c\tau + d)^{1/2}=:\epsilon J_{1/2}(\gamma, \tau)$ with $\epsilon=\pm 1$, 
the above law is written by
\begin{align}
    (\gamma_1, \epsilon_1 J_{1/2}(\gamma_1, \tau))(\gamma_2, \epsilon_2 J_{1/2}(\gamma_2, \tau))
    = (\gamma_1 \gamma_2, \epsilon_1 \epsilon_2 \zeta_{1/2}(\gamma_1, \gamma_2) J_{1/2}(\gamma_1\gamma_2,\tau)),    
\end{align}
where $\zeta_{1/2}(\gamma_1, \gamma_2)=\{ +1, -1\}$ (for more details, see, e.g., Appendix A of Ref. \cite{Liu:2020msy}). 
For practical purposes, we present some explicit forms of $\zeta_{1/2}(\gamma_1, \gamma_2)$:
\begin{align}
    \zeta_{1/2}(\gamma, T) &= \zeta_{1/2}(T, \gamma) = 1,
    \nonumber\\
    \zeta_{1/2}(\gamma, S) &= 
    \left\{
    \begin{array}{l}
         -1,\qquad (c<0, d\leq 0)\\
         1,\qquad ({\rm others})
    \end{array}
    \right.
    ,\qquad
    \zeta_{1/2}(S, \gamma) = 
    \left\{
    \begin{array}{l}
         -1,\qquad (a\leq 0, b< 0)\\
         1,\qquad ({\rm others})
    \end{array}
    \right.
    .    
\end{align}

\medskip

The generators of $Mp(2,\mathbb{Z})$ are written in terms of $SL(2,\mathbb{Z})$ generators $S$ and $T$:
\begin{align}
    \widetilde{S} = (S, -\sqrt{-\tau}) 
    = \left( \left( 
    \begin{array}{cc}
        0 & 1 \\
        -1 & 0
    \end{array}
    \right),
    -\sqrt{-\tau} \right),
    \qquad
    \Tilde{T} = (T, 1) 
    = \left( \left( 
    \begin{array}{cc}
        1 & 1 \\
        0 & 1
    \end{array}
    \right),
    1 \right),
\end{align}
satisfying 
\begin{align}
    \widetilde{S}^2 = \Tilde{R},\qquad
    (\widetilde{S}\Tilde{T})^3 = \Tilde{R}^4 = 1,\qquad
    \Tilde{T}\Tilde{R} = \Tilde{R}\Tilde{T}.
\end{align}
Here, $\Tilde{R}$ and $\widetilde{S}\Tilde{T}$ are of the form
\begin{align}
    \Tilde{R} &:= \widetilde{S}^2 = (S^2,-i) 
    = \left( \left( 
    \begin{array}{cc}
        -1 & 0 \\
        0 & -1
    \end{array}
    \right),
    -i \right),
\nonumber\\
    \widetilde{S}\Tilde{T} &= (ST, -\sqrt{-\tau-1}) 
    = \left( \left( 
    \begin{array}{cc}
        0 & 1 \\
        -1 & -1
    \end{array}
    \right),
    -\sqrt{-\tau-1} \right).
\end{align}
As presented above, there exist two elements $(\gamma, \pm J_{1/2}(\gamma, \tau))$ of $Mp(2,\mathbb{Z})$ for each $\gamma \in SL(2, \mathbb{Z})$. Furthermore, $\tilde{R}^2$ leads to $\tilde{R}^2(\gamma, J_{1/2}(\gamma, \tau))= (\gamma, -J_{1/2}(\gamma, \tau))$. 
Thus, the metaplectic group is a twofold cover group of $SL(2, \mathbb{Z})$, that is, $SL(2,\mathbb{Z})\simeq Mp(2,\mathbb{Z})/\mathbb{Z}_2$ with $\mathbb{Z}_2=\{1, \tilde{R}^2\}$. 
It was known that one can define the finite modular groups in both $SL(2,\mathbb{Z})$ and $Mp(2,\mathbb{Z})$. 

\medskip

Let us rewrite $SL(2, \mathbb{Z})$, its quotient group and metaplectic group by
\begin{align}
\Gamma := SL(2,\mathbb{Z}),\qquad
\Bar{\Gamma}:= \Gamma/\{\pm \mathbb{I}\},\qquad
\tilde{\Gamma} := Mp(2,\mathbb{Z}),
\end{align}
respectively. Note that the complex structure moduli space of the torus $\tau$ is governed by $\Bar{\Gamma}$ due to the fact that $\tau$ is invariant under $S^2$. 
By introducing the principal congruence subgroups:
\begin{align}
\Gamma(N) &=
\left\{
    \left(
    \begin{array}{cc}
        a & b \\
        c & d
    \end{array}
    \right)
    \in \Gamma\,
    \Biggl|
    a\equiv d \equiv 1,\qquad b\equiv c\equiv 0\,({\rm mod}\,N)
    \right\}
    ,
\nonumber\\ 
\bar{\Gamma}(N) &= \Gamma(N)/\{ \pm \mathbb{I}\},
\nonumber\\
\tilde{\Gamma}(4N) &=
\left\{
\tilde{\gamma}= (\gamma, v(\gamma)J_{1/2}(\gamma, \tau))
    \bigl|
    \,
\gamma \in \Gamma (4N)
    \right\}
    ,
\end{align}
with $v(\gamma) = \left( \frac{c}{d}\right)$ being the Kronecker symbol, 
one can define the finite modular groups:
\begin{align}
    \Gamma_N := \bar{\Gamma}/\bar{\Gamma}(N) = \langle S, T | S^4 = (ST)^3 = T^N=1\rangle,
\end{align}
where $\Gamma_{2,3,4,5}$ correspond to $S_3, A_4, S_4, A_5$ discrete groups, respectively.\footnote{To construct the finite modular groups with $N>5$, we require an additional relation.} 
In addition, the finite metaplectic modular groups are given by
\begin{align}
    \tilde{\Gamma}_{4N} := \tilde{\Gamma}/\tilde{\Gamma}(4N),
\end{align}
where the generators satisfy\footnote{Hereafter, $\widetilde{S}$ and $\widetilde{T}$ denotes the generators of the finite metaplectic modular groups.}
\begin{align}
    \widetilde{S}^2 = \Tilde{R},\qquad
    (\widetilde{S}\Tilde{T})^3 = \Tilde{R}^4 = 1,\qquad
    \Tilde{T}\Tilde{R} = \Tilde{R}\Tilde{T},\qquad
    \tilde{T}^{4N}=\mathbb{I},
\label{eq:meta_generator}
\end{align}
and additional relations are required to ensure the finiteness for $N>1$, e.g., 
\begin{align}
&\widetilde{S}^5\tilde{T}^6\widetilde{S}\tilde{T}^4\widetilde{S}\tilde{T}^2\widetilde{S}\tilde{T}^4 = \mathbb{I},\qquad    ({\rm for}\,N=2),
\label{eq:meta_generatorN=2}
\end{align}
for $\tilde{\Gamma}_{4N=8}$ of order 768 ([768, 1085324] in GAP system \cite{GAP4}), 
\begin{align}
&\widetilde{S}\tilde{T}^3\widetilde{S}\tilde{T}^{-2}\widetilde{S}^{-1}\tilde{T}\widetilde{S}\tilde{T}^{-3}\widetilde{S}^{-1}\tilde{T}^2\widetilde{S}^{-1}\tilde{T}^{-1} = \mathbb{I},\qquad    ({\rm for}\,N=3),
\label{eq:meta_generatorN=3}
\end{align}
for $\tilde{\Gamma}_{4N=12}$ of order 2304, respectively. 
Under the finite modular groups, modular forms of the modular weight $k/2$ and level $4N$ transform as
\begin{align}
    f_{\tilde{\alpha}}(\tilde{\gamma}\tau) = \varphi(\gamma, \tau)^k \rho_{\bf r}(\tilde{\gamma})_{\tilde{\alpha}\tilde{\beta}} f_{\tilde{\beta}}(\tau),
\end{align}
where $\rho_{\bf r}(\tilde{\gamma})_{\tilde{\alpha}\tilde{\beta}}$ denotes an irreducible representation matrix in $\tilde{\Gamma}_{4N}$.

\subsection{$T^2$ with magnetic fluxes}
\label{sec:T2}

As discussed in Secs. \ref{sec:generation1} and \ref{sec:generation2}, the magnetic 
fluxes generate the semi-realistic MSSM-like models with 3 generations of quarks and 
leptons. 
In the following, we address the flavor structure of chiral zero modes with an emphasis 
on the transformations of these wavefunctions under the modular symmetry. 
It was known that the magnetic fluxes on extra-dimensional spaces induce the degenerate chiral zero-modes, 
which are counted by the index theorem. 

\medskip

For concreteness, let us begin with the six-dimensional (6D) Super Yang-Mills theory on $T^2$.  
The Kaluza-Klein reduction of 6D Majorana-Weyl spinor $\lambda$ is given by
\begin{align}
    \lambda(x,z) = \sum_n \phi_n (x) \otimes \psi_n(z),
\end{align}
with $\psi_n(z)$ denotes the $n$-th excited mode of two-dimensional (2D) Weyl spinors on $T^2$. 
In particular, we focus on zero-mode wavefunctions $\psi(z)$:\footnote{Here and in what follows, we omit the zero-mode 
index.}
\begin{align}
    \psi(z) = 
    \left(
    \begin{array}{c}
         \psi_+(z)\\
         \psi_-(z) 
    \end{array}
    \right),
\end{align}
where $\psi_+$ and $\psi_-$ denote the positive and negative chirality modes on $T^2$. 
The $U(1)$ magnetic flux is given by
\begin{align}
    F = \frac{i\pi M}{{\rm Im}z} dz \wedge d\bar{z},
\end{align}
obtained by the corresponding vector potential
\begin{align}
A = \frac{\pi M}{{\rm Im}z}{\rm Im}((\bar{z}+\bar{\zeta})dz),
\end{align}
with $\zeta$ being a Wilson line phase. 
Note that the boundary conditions of the gauge field as well as the 2D Weyl spinors are 
respectively chosen as
\begin{align}
    A(z+1, \bar{z}+1) &= A(z, \bar{z}) + d\chi_1(z,\bar{z}),\quad    
    A(z+\tau, \bar{z}+\bar{\tau}) = A(z, \bar{z}) + d\chi_2(z,\bar{z})
    \nonumber\\
    \psi(z+1) &= e^{i\chi_1}\psi(z),\qquad \qquad \qquad \qquad\,
    \psi(z+\tau) = e^{i\chi_2}\psi(z),    
    \label{eq:BCs}
\end{align}
with
\begin{align}
    \chi_1 = \frac{\pi M}{{\rm Im}\tau}{\rm Im}(z+\zeta),\qquad
    \chi_2 = \frac{\pi M}{{\rm Im}\tau}{\rm Im}\bigl[ \bar{\tau}(z+\zeta)\bigl].
\end{align}
By solving the Dirac equation for the massless mode with $U(1)$ charge $q=1$, 
we find $|M|$ degenerate zero-mode solutions; $\psi_+ (z)$ for $M>0$ and 
$\psi_- (z)$ for $M<0$. Specifically, $|M|$ degenerate zero-mode 
wavefunctions are written in terms of Jacobi theta function $\vartheta$ 
and the torus area ${\cal A}$~\cite{Cremades:2004wa}:
\begin{align}
    \psi_{\pm}^{\talpha, |M|}(z,\tau) = \left( \frac{|M|}{{\cal A}^2}\right)^{1/4} e^{i\pi |M| (z+\zeta)\frac{{\rm Im}(z+\zeta)}{{\rm Im}\tau}}
      \vartheta \begin{bmatrix}
    \frac{\talpha}{M} \\
    0 \\
  \end{bmatrix}
  (|M|z,|M|\tau), 
  \label{eq:psi} 
\end{align}
with $\talpha=0,1,...,|M|-1$ and
\begin{align}
  \vartheta
  \begin{bmatrix}
    a \\ b \\
  \end{bmatrix}
  (\nu,\tau)
  := \sum_{\ell\in\mathbb{Z}} e^{\pi i(a+\ell)^2\tau} e^{2\pi i(a+\ell)(\nu+b)}.
\label{eq:theta}
\end{align}
Here, the normalization of the wavefunctions is fixed as\footnote{In what follows, we omit the index of chirality.}
\begin{align}
  &\int d^2z ~\psi^{\tilde{\alpha},M}(z,\tau) \left(\psi^{\tilde{\beta},M} (z,\tau)\right)^* = (2{\rm Im}\tau)^{-1/2}\delta_{\talpha,\tbeta}. 
  \label{eq:Normalization}
\end{align}
The Yukawa couplings of chiral zero-modes are obtained by integrals of three wavefunctions:
\begin{align}
    Y^{\talpha \tbeta \tgamma} = \int_{T^2} d^2z\, \psi^{\tilde{\alpha},|M|}(z,\tau)\psi^{\tilde{\beta},|M'|}(z,\tau)(\psi^{\tilde{\gamma},|M|+|M'|}(z,\tau))^\ast.
    \label{eq:Yukawa_general}
\end{align}

\medskip

Remarkably, these wavefunctions show non-trivial transformations under the modular symmetry \cite{Cremades:2004wa,Kobayashi:2018rad,Kobayashi:2018bff,
Ohki:2020bpo,Kikuchi:2020frp,Kikuchi:2020nxn,
Kikuchi:2021ogn,Almumin:2021fbk}. 
Indeed, when $|M|=$ even, under $S$ and $T$ transformations of the modular symmetry:
\begin{align}
S&:~\tau \to -1/\tau, \qquad z \to -z/\tau,
\nonumber\\
T&:~\tau \to \tau+1, \qquad z \to z,
\end{align}
the zero-mode wavefunctions respectively transform 
\begin{align}
\psi^{\talpha,|M|}(z,\tau) &\to \psi^{\talpha,|M|}\left(-\frac{z}{\tau},-\frac{1}{\tau}\right) = (-\tau)^{1/2} \sum_{\tbeta=0}^{|M|-1} \frac{1}{\sqrt{|M|}}e^{i \pi /4}e^{2\pi i\,\frac{\talpha \tbeta}{|M|}} \psi^{\tbeta,|M|}(z,\tau),
\nonumber\\
\psi^{\talpha,|M|}(z,\tau) &\to \psi^{\talpha,|M|}(z,\tau+1) = e^{i\pi \frac{\talpha^2}{|M|}} \psi^{\talpha,|M|}(z,\tau),
\label{eq:wavemod_even}
\end{align}
indicating the wavefunctions with the modular weight $1/2$.\footnote{The modular weights in Type IIB magnetized D-brane models and the T-dual Type IIA intersecting D-brane models are revisited in Ref. \cite{Kikuchi:2023clx}.}
Note that the Wilson line $\zeta$ transform as $\zeta \rightarrow \zeta/(c\tau +d)$ as in the coordinate $z$. 
By taking $M=2M'$ with $M'\in \mathbb{Z}$, they are  rewritten in the context of $\tilde{\Gamma}_{2|M|=4|M'|}$  
\begin{align}
    \psi^{\tilde{\alpha}, |M|}(\tilde{\gamma}(z, \tau)) = \varphi(\gamma, \tau) \sum_{\tilde{\beta}=0}^{2|M'|-1} \rho (\tilde{\gamma})_{\tilde{\alpha}\tilde{\beta}} \psi^{\tilde{\beta}, |M|}(z,\tau),
\end{align}
with $\tilde{\alpha},\tilde{\beta}=0,1,...,2|M'|-1$ and
\begin{align}
    \rho(\widetilde{S})_{\tilde{\alpha}\tilde{\beta}} &=-\frac{1}{\sqrt{|M|}}e^{i \pi /4}e^{2\pi i\,\frac{\tilde{\alpha}\tilde{\beta}}{|M|}},
    \label{eq:metaS_old}
    \\
    \rho(\widetilde{T})_{\tilde{\alpha}\tilde{\beta}} &= e^{2\pi i\,\frac{\tilde{\beta}^2}{|M|}} \delta_{\tilde{\alpha}, \tilde{\beta}}.
    \label{eq:metaT_old}
\end{align}
Indeed, the representation matrix $\rho(\widetilde{\gamma})$ is unitary and satisfies 
\begin{align}
    \rho(\widetilde{S})^2 = \rho(\widetilde{R}),\quad
    (\rho(\widetilde{S})\rho(\widetilde{T}))^3 = \rho(\widetilde{R}^4) = 1,\quad
    \rho(\widetilde{T})\rho(\widetilde{R}) = \rho(\widetilde{R})\rho(\widetilde{T}),\quad
    \rho(\widetilde{T})^{4|M'|}=\mathbb{I}.
\end{align}

It was known that the boundary conditions of the fermions in Eq. (\ref{eq:BCs}) and the $T$ transformation 
are consistent with each other only if $M$ is even. The $S$ transformation is consistent with 
the boundary conditions. 
However, the existence of Wilson line modifies the boundary condition as well as the modular transformation \cite{Kikuchi:2021ogn}. 
Taking into account the modular transformation of the Wilson line $\zeta$ in the case of $M=$ odd \footnote{We 
discuss the modular transformation of the Wilson lines, but they are related to that of the Scherk-Schwarz phases \cite{Abe:2013bca}. See, Appendix of Ref. \cite{Kikuchi:2021ogn}, for more details about the modular transformations of Scherk-Schwarz phases and Wilson lines.}, 
\begin{align}
    M\zeta \rightarrow M\left(\zeta+\frac{1}{2}\right),
\end{align}
the wavefunction transforms under the $T$ transformation:
\begin{align}
    \psi^{\talpha, |M|} \left( z+ \zeta + \frac{1}{2}, \tau +1\right) &= e^{i\pi|M|\frac{{\rm Im}\,(z+\zeta)}{2{\rm Im}\,\tau}}e^{i\pi j \left( \frac{\talpha}{|M|}+1\right)} \psi^{\talpha, |M|}(z+\zeta, \tau)
    \nonumber\\
    &=\sum_{\tbeta}e^{i\pi|M|\frac{{\rm Im}\,(z+\zeta)}{2{\rm Im}\,\tau}}\rho(\widetilde{T})_{\talpha \tbeta} \psi^{\tbeta, |M|}(z+\zeta, \tau),
\label{eq:wavemod_odd}
\end{align}
with
\begin{align}
    \rho(\widetilde{T})_{\talpha \tbeta} = e^{i\pi \talpha \left( \frac{\talpha}{|M|}+1\right)}\delta_{\talpha,\tbeta}.
    \label{eq:metaT_new}
\end{align}
Note that the authors of Ref. \cite{Almumin:2021fbk} proposed that this expression holds for vanishing Wilson lines even in the case of odd units of magnetic flux $M$. Recall that the exponential factor $e^{i\pi|M|\frac{{\rm Im}\,(z+\zeta)}{2{\rm Im}\,\tau}}$ can be canceled in 
the Yukawa coupling due to the $U(1)$ gauge invariance as argued in Ref. \cite{Almumin:2021fbk}. 
Thus, $T$-transformed wavefunction with odd $M$ cannot be expanded in terms of the original wavefunction, but it will be possible to be written in the different coordinate $z+1/2$. 
Since this statement is also true for even units of $M$, we adopt the $T$ transformation of wavefunction is described by Eq. (\ref{eq:wavemod_odd}) for a general $M$. 
The $S$ transformation is still given by Eq. (\ref{eq:metaS_old}) with odd units of $M$. 

\medskip

The modular transformations also act on the 4D fields. 
When the 4D ${\cal N}=1$ SUSY is preserved, the 4D Lagrangian is written in terms of K\"ahler potential 
and superpotential. 
The K\"ahler potential and the superpotential of matter fields are 
derived from the dimensional reduction of 6D Super Yang-Mills theory:
\begin{align}
    K &= \frac{|\Phi|^2}{(i(\bar{\tau}-\tau))^{1/2}},
\nonumber\\
    W &= Y_{\talpha \tbeta \tgamma} \Phi^{\talpha, I_{ab}}\Phi^{\tbeta, I_{ca}}\Phi^{\tgamma, I_{cb}},
\end{align}
where $\{I_{ab}, I_{ca}, I_{cb}\}$ denote the generation number counted by the index theorem, 
corresponding to one of the torus in Eq. (\ref{eq:Index}). 
It satisfies $I_{ab}+I_{bc}+I_{ca}=0$ to preserve the $U(1)$ gauge symmetry. 
Then, the modular transformations of matter superfield are given by
\begin{align}
    \Phi^{\talpha, I_{ab}} \rightarrow \varphi(\gamma, \tau)^{-1} \left(\rho(\tilde{\gamma})\right)_{\talpha \tbeta}^{-1} \Phi^{\tbeta,I_{ab}},
\end{align}
where explicit forms of $\rho(\tilde{\gamma})$ are given in Eqs. (\ref{eq:metaS_old}) and (\ref{eq:metaT_new}) by replacing $M$ with $I_{ab}$. 
In addition, it was known that the holomorphic Yukawa couplings (\ref{eq:Yukawa_general}) are also described by Jacobi theta function \cite{Cremades:2004wa}:
\begin{align}
Y_{\talpha \tbeta \tgamma} \simeq \sigma_{abc} \left( \frac{2{\rm Im}\,\tau}{{\cal A}^2}\right)^{1/4}
\biggl|\frac{I_{ab}I_{ca}}{I_{bc}} \biggl|^{1/4} 
e^{H(\Tilde{\zeta},\tau)/2} 
  \vartheta
  \begin{bmatrix}
    \delta_{\talpha \tbeta \tgamma} \\ 0 \\
  \end{bmatrix}
  (\Tilde{\zeta}, \tau |I_{ab}I_{bc}I_{ca}|),
\label{eq:Yukawa}
\end{align}
with 
\begin{align}
\sigma_{abc} &= {\rm sign}(I_{ab}I_{bc}I_{ca}),
\nonumber\\
    \delta_{\talpha \tbeta \tgamma} &= \frac{\talpha}{I_{ab}} + \frac{\tbeta}{I_{ca}} + \frac{\tgamma}{I_{bc}},
    \nonumber\\
    \Tilde{\zeta} &= I_{ab}\tilde{\zeta}_c + I_{bc}\tilde{\zeta}_a + I_{ca}\tilde{\zeta}_b,
    \nonumber\\
     H(\Tilde{\zeta}, \tau) &= 2\pi i |I_{ab}I_{bc}I_{ca}|^{-1} \frac{\Tilde{\zeta}\cdot {\rm Im}\,\Tilde{\zeta}}{{\rm Im}\,\tau},
\end{align}
where $\Tilde{\zeta}_a = n_a\zeta_a/m_a$ denotes the redefined Wilson lines and we omit the 6D gauge coupling in the 
above expression. 
Since the Yukawa couplings are described by the half-integer modular form, 
the Yukawa couplings belong to ${\bf r}$ representation of $\tilde{\Gamma}_{4N}$ 
whose transformation is of the form\footnote{Note that the Jacobi theta function itself behaves a proper modular form satisfying
$    (Y_{\bf r})_{\Tilde{\alpha}}(\tau) \rightarrow (Y_{\bf r})_{\Tilde{\alpha}}(\Tilde{\gamma}\tau) 
    = \varphi(\gamma, \tau) \rho_{\bf r}(\Tilde{\gamma})_{\Tilde{\alpha}\Tilde{\beta}}(Y_{\bf r})_{\Tilde{\beta}}(\tau)$.}:
\begin{align}
    (Y_{\bf r})_{\Tilde{\alpha}}(\tau) \rightarrow (Y_{\bf r})_{\Tilde{\alpha}}(\widetilde{\gamma}\tau) 
    = \rho_{\bf r}(\widetilde{\gamma})_{\Tilde{\alpha}\Tilde{\beta}}(Y_{\bf r})_{\Tilde{\beta}}(\tau).
\end{align}
Recalling the condition $I_{ab}+I_{bc}+I_{ca}=0$, the Yukawa terms are invariant under the following $U(1)$ symmetry:
\begin{align}
    \Phi^{\talpha, I_{ab}} \rightarrow e^{iq\alpha I_{ab}}\Phi^{\talpha,I_{ab}},
\end{align}
with $q$ being the $U(1)$ charge of $\Phi^{\talpha,I_{ab}}$. 
Thus, we redefine the $\widetilde{S}$ transformation of matter fields following Ref. \cite{Almumin:2021fbk}:
\begin{align}
\rho(\widetilde{S})_{\tilde{\alpha} \tilde{\beta}} = -\frac{1}{\sqrt{|I_{ab}|}}e^{i \pi \frac{3|I_{ab}|+1}{4}}e^{2\pi i\,\frac{\tilde{\alpha}\tilde{\beta}}{|I_{ab}|}}.
\label{eq:metaS_new}
\end{align}
Although we add $e^{3i\pi I_{ab}/4}$ in the $S$ transformation, it is still the unitary representation matrix. 
Such a redefinition will be convenient to discuss the metaplectic modular symmetry as will be shown later. 
In this way, the $T^2$ compactifications with magnetic background fluxes lead to the metaplectic modular flavor symmetries. 
Before going into details about the relation between the metaplectic modular flavor symmetries, the traditional flavor and CP symmetries, 
we will discuss the metaplectic modular symmetry on $T^2/\mathbb{Z}_2$ background.

\subsection{$T^2/\mathbb{Z}_2$ with magnetic fluxes}
\label{sec:T2Z2}

On the $T^2/\mathbb{Z}_2$ orbifold, the wavefunctions of $\mathbb{Z}_2$-even and -odd modes are given by the linear combination of these on $T^2$ as mentioned in Sec. \ref{sec:generation1}. 
The explicit forms are given by \cite{Abe:2008fi}
\begin{align}
\begin{split}
    \psi_{\rm even}^{\talpha,|M|} &= {\cal N}_{\talpha} \sum_{\tbeta=0}^{|M|-1}
    \left( \delta_{\talpha, \tbeta} + \delta_{|M|-\talpha, \tbeta}\right) \psi^{\tbeta, |M|},
    \\
    \psi_{\rm odd}^{\talpha,|M|} &= {\cal N}_j \sum_{\tbeta=0}^{|M|-1}
    \left( \delta_{\talpha, \tbeta} - \delta_{|M|-\talpha, \tbeta}\right) \psi^{\tbeta, |M|},
\end{split}
\label{eq:wave_evenodd}
\end{align}
with
\begin{align}
    {\cal N}_{\talpha} =
    \left\{
    \begin{array}{l}
         1 \qquad (\talpha=0)  \\
         \frac{1}{2}\qquad \left(\talpha=\frac{|M|}{2}\right)\\
         \frac{1}{\sqrt{2}}\qquad ({\rm others})\\
    \end{array}
    \right.
    .
\end{align}
The modular transformations are extracted from the matter wavefunctions on $T^2$ (\ref{eq:metaT_new}) and (\ref{eq:metaS_new}):
\begin{align}
\rho(\widetilde{S})_{\tilde{\alpha}\tilde{\beta}} &= -\frac{2}{\sqrt{|M|}}e^{i \pi \frac{3|M|+1}{4}}\,\cos \left(\frac{2\pi \tilde{\alpha}\tilde{\beta}}{|M|}\right),
\label{eq:metaS_Z2even}
\\
    \rho(\widetilde{T})_{\talpha \tbeta} &= e^{i\pi \talpha \left( \frac{\talpha}{|M|}+1\right)}\delta_{\talpha,\tbeta},
    \label{eq:metaT_Z2even}
\end{align}
for $\mathbb{Z}_2$-even mode with $\tilde{\alpha},\tilde{\beta} = 0,1,...,I_{\rm even}$ and
\begin{align}
\rho(\widetilde{S})_{\talpha\tbeta} &= -\frac{2i}{\sqrt{|M|}}e^{i \pi \frac{3|M|+1}{4}}\,\sin \left(\frac{2\pi \talpha\tbeta}{|M|}\right),
\label{eq:metaS_Z2odd}
\\
    \rho(\widetilde{T})_{\talpha\tbeta} &= e^{i\pi \talpha \left( \frac{\talpha}{|M|}+1\right)}\delta_{\talpha,\tbeta},
    \label{eq:metaT_Z2odd}
\end{align}
for $\mathbb{Z}_2$-odd mode with $\tilde{\alpha},\tilde{\beta} = 0,1,...,I_{\rm odd}$. 
Here, $I_{\rm even}$ and $I_{\rm odd}$ are defined in Eq. (\ref{eq:index_evenodd}). 
In contrast to the analysis of Ref. \cite{Kikuchi:2020frp}, we added 
extra phase factors as argued in the previous section. 
Since Yukawa couplings on $T^2/\mathbb{Z}_2$ are described by those on $T^2$ \cite{Kikuchi:2021yog}:
\begin{align}
Y_{\talpha'\tbeta'\tgamma'}\biggl|_{T^2/\mathbb{Z}_2} = \sum {\cal O}^{\talpha'\talpha,|I_{ab}|}{\cal O}^{\tbeta'\tbeta,|I_{bc|}}{\cal O}^{\tgamma'\tgamma,|I_{ca}|}
    Y_{\talpha\tbeta\tgamma}\biggl|_{T^2},
\end{align}
with
\begin{align}
    {\cal O}_m^{\talpha\tbeta,|M|}={\cal N}_{\talpha} \left(\delta_{\talpha,\tbeta} + (-1)^m\delta_{\talpha,|M|-\tbeta}\right), 
\end{align}
where $m=0$ and $m=1$ respectively correspond to $\mathbb{Z}_2$-even and -odd modes, 
they transform under the metaplectic modular symmetry as in the matter fields. 

\medskip

We find that the unitary matrices (\ref{eq:metaT_Z2even}) and (\ref{eq:metaT_Z2odd}) obey the required relations in the metaplectic 
modular symmetry (\ref{eq:meta_generator}). 
Specifically, for even $|M|$ and odd $|M|$,  modular transformations of both the $\mathbb{Z}_2$-even and -odd modes are described by $\Tilde{\Gamma}_{2|M|}$ and $\Tilde{\Gamma}_{4|M|}$, respectively, 
We checked the additional relations (\ref{eq:meta_generatorN=2}) and (\ref{eq:meta_generatorN=3}) for $\tilde{\Gamma}_{8}$ and $\tilde{\Gamma}_{12}$, respectively. However, we have not checked 
these additional relations with $\Tilde{\Gamma}_{4N}$ ($N\geq 4$) which 
will be reported elsewhere.  
In the next section, we derive such 6D bottom-up models from 10D Type IIB magnetized D-brane models with stabilized moduli. 
In particular, we discuss the metaplectic modular flavor symmetries together with 
traditional flavor and CP symmetries in the framework of eclectic symmetry.

\subsection{Generalized CP}
\label{sec:CP}

We first discuss the unification of the metaplectic modular symmetry and 4D CP symmetry, which was discussed in $T^2$ background \cite{Kikuchi:2020frp}. 
It was known that the 4D CP and 6D orientation reversing are embedded into the 10D proper Lorentz symmetry \cite{Dine:1992ya,Choi:1992xp}.\footnote{The CP symmetry is discussed in the context of Type IIB string theory, e.g., Refs.\cite{Kobayashi:2020uaj,Ishiguro:2020nuf}.} 
Since the 6D orientation reversing is realized by $z_i \rightarrow - \Bar{z}_i$ for the coordinates of $(T^2)_i$, the torus modulus transform under the CP symmetry $\tau_i \rightarrow - \bar{\tau}_i$. 
Note that such a transformation leads to the negative determinant in the transformation of 6D space. 
In the following, we focus on the CP transformation on $T^2$ and $T^2/\mathbb{Z}_2$.

\medskip

Recalling that the modulus is defined by the ratio of the lattice vectors of the torus $\tau_i = e_2/e_1$, the CP transformation is realized by
\begin{align}
    \begin{pmatrix}
        e_2\\
        e_1\\
    \end{pmatrix}
    \xrightarrow{CP}
    \begin{pmatrix}
        1 & 0\\
        0 & -1 
    \end{pmatrix}
        \begin{pmatrix}
        \bar{e}_2\\
        \bar{e}_1\\
    \end{pmatrix}
    \label{eq:CP}
    .
\end{align}
Thus, the CP matrix is not the element of $SL(2,\mathbb{Z})$, and 
the CP transformation enlarges the $SL(2,\mathbb{Z})$ modular group 
to $GL(2,\mathbb{Z})\simeq SL(2,\mathbb{Z}) \rtimes \mathbb{Z}_2^{\rm CP}$ \cite{Baur:2019kwi,Novichkov:2019sqv,Baur:2019iai}.\footnote{In the case of the multi moduli, such a CP transformation also enlarges the symplectic modular group to $GSp(2g,\mathbb{Z})\simeq Sp(2g, \mathbb{Z})\rtimes \mathbb{Z}_2^{\rm CP}$ \cite{Ishiguro:2020nuf}.}
Indeed, the CP matrix (\ref{eq:CP}) and the generators of $SL(2,\mathbb{Z})$ 
satisfy
\begin{align}
    CP^2 = 1, \quad
    (CP)S(CP)^{-1} = S^{-1},\quad
    (CP)T(CP)^{-1} = T^{-1},    
\end{align}
with 
\begin{align}
    S =
    \begin{pmatrix}
        0 & 1 \\
        -1 & 0 \\
    \end{pmatrix}
    ,
    \qquad
    T =
    \begin{pmatrix}
        1 & 1 \\
        0 & 1 \\
    \end{pmatrix}
    .
\end{align}
As seen in Sec. \ref{sec:meta}, the metaplectic modular group is 
defined for $\gamma \in SL(2,\mathbb{Z})$. 
For $\gamma^\ast \in GL(2,\mathbb{Z})\simeq SL(2,\mathbb{Z}) \rtimes \mathbb{Z}_2^{\rm CP}$, the modulus $\tau$ as well as the automorphy factor $\varphi(\gamma^\ast, \tau)$ transform as follows (see Ref. \cite{Kikuchi:2020frp}):
\begin{align}
    \tau &\xrightarrow{\gamma^\ast}
    \left\{
    \begin{array}{l}
        \frac{a\tau +b}{c\tau +d}\qquad ({\rm det}(\gamma^\ast)=1)\\
        \frac{a\bar{\tau} +b}{c\bar{\tau} +d}\qquad ({\rm det}(\gamma^\ast)=-1)\\
    \end{array}
    \right.
    ,
    \nonumber\\
    \varphi(\gamma^\ast, \tau)&= 
    \left\{
    \begin{array}{l}
        \pm (c\tau +d)^{1/2}\qquad ({\rm det}(\gamma^\ast)=1)\\
        \pm (c\bar{\tau} +d)^{1/2}\qquad ({\rm det}(\gamma^\ast)=-1)\\
    \end{array}
    \right.
    .
\end{align}
The multiplication law in the context of metaplectic modular symmetry is defined as
\begin{align}
    (\gamma_1^\ast, \varphi(\gamma_1^\ast, \tau))(\gamma_2^\ast, \varphi(\gamma_2^\ast, \tau))
    = (\gamma_1^\ast \gamma_2^\ast, \varphi(\gamma_1^\ast, \gamma_2^\ast \tau)\varphi(\gamma_2^\ast,\tau)).
\end{align}
When we redefine $\varphi(\gamma^\ast, \tau) = \pm (c\tau + d)^{1/2}=:\epsilon J_{1/2}(\gamma^\ast, \tau)$ with $\epsilon=\pm 1$, 
the above law is written by
\begin{align}
    (\gamma_1^\ast, \epsilon_1 J_{1/2}(\gamma_1^\ast, \tau))(\gamma_2^\ast, \epsilon_2 J_{1/2}(\gamma_2, \tau))
    = (\gamma_1^\ast \gamma_2^\ast, \epsilon_1 \epsilon_2 \zeta_{1/2}^\ast(\gamma_1^\ast, \gamma_2^\ast) J_{1/2}(\gamma_1^\ast\gamma_2^\ast,\tau)),    
\end{align}
where the two-cocyle $\zeta_{1/2}^\ast(\gamma_1^\ast, \gamma_2^\ast)$ is defined as
\begin{align}
    \zeta_{1/2}^\ast(\gamma_1^\ast, \gamma_2^\ast) = ({\rm det}\gamma_1^\ast, {\rm det}\gamma_2^\ast)_H 
    \left(\frac{\chi(\gamma_1^\ast\gamma_2^\ast)}{\chi(\gamma_1^\ast)}, \frac{\chi(\gamma_1^\ast\gamma_2^\ast)}{\chi(\gamma_2^\ast){\rm det}\gamma_1^\ast} \right)_H
    \frac{s(\gamma_1^\ast)s(\gamma_2^\ast)}{s(\gamma_1^\ast \gamma_2^\ast)}.
\end{align}
Here, we define
\begin{align}
    s(\gamma^\ast) = 
    \left\{
    \begin{array}{c}
            1,\,\,\,\,\,\,\qquad (c\neq 0)\\
        {\rm sign}(d),\quad(c=0)\\
    \end{array}
    \right.,
    \qquad
    \chi(\gamma^\ast)
    =
    \left\{
    \begin{array}{c}
            c,\qquad (c\neq 0)\\
            d,\qquad (c=0)\\
    \end{array}
    \right.
    ,
\end{align}
and introduce the Hilbert symbol:
\begin{align}
    (x,y)_H =
    \left\{
    \begin{array}{l}
            -1,\qquad (x< 0\,\,{\rm and}\,\,y<0)\\
            +1,\qquad ({\rm others})\\
    \end{array}
    \right.
    .
\end{align}

\medskip

Since the CP transformation of matter wavefunctions on $T^2$ is given by
\begin{align}
    \psi^{\talpha, M}\xrightarrow{CP} \overline{\psi^{\talpha,|M|}(z,\tau)},
    \label{eq:CP_wave}
\end{align}
corresponding to the basis of canonical CP transformation\footnote{Note that the CP transformation flips the sign of the magnetic flux, that is, $M\rightarrow -M$.}, 
it allows us to define the CP transformation in the framework of 
metaplectic modular symmetry:
\begin{align}
    \widetilde{CP}=[CP, i]
    = \left( \left( 
    \begin{array}{cc}
        1 & 0 \\
        0 & -1
    \end{array}
    \right),
    i \right).
\end{align}
Thus, Eq. (\ref{eq:CP_wave}) is rewritten as
\begin{align}
    \psi^{\talpha, M} \rightarrow \varphi(\widetilde{CP}, \tau) \rho(\widetilde{CP})_{\talpha\tbeta}\overline{\psi^{\tbeta,|M|}(z,\tau)}
\end{align}
with 
\begin{align}
\varphi(\widetilde{CP},\tau)= i,
\qquad \rho(\widetilde{CP})_{\talpha\tbeta} = -i \delta_{\talpha,\tbeta}.
\end{align}

\medskip

Remarkably, the CP transformation does not commute with the metaplectic modular transformations. 
When we consider the following chain:
\begin{align}
\psi \xrightarrow{\widetilde{CP}} 
&\,\rho(\widetilde{CP})\overline{\psi}  
\xrightarrow{\widetilde{\gamma}} 
\varphi(\widetilde{\gamma},\tau)^\ast \rho(\widetilde{CP}) \rho(\widetilde{\gamma})^\ast \overline{\psi}
\nonumber\\
\xrightarrow{\widetilde{CP}^{-1}} 
&\,\varphi(\widetilde{\gamma},\tau) \rho(\widetilde{CP}) \rho(\widetilde{\gamma})^\ast \rho(\widetilde{CP})^{-1} \psi,
\end{align}
for $\widetilde{\gamma} \in Mp(2,\mathbb{Z})$, 
we find that 
\begin{align}
    \rho(\widetilde{CP})\rho(\widetilde{S})^\ast \rho(\widetilde{CP})^{-1}
    &= \rho(\widetilde{S})^{-1},
    \nonumber\\
    \rho(\widetilde{CP})\rho(\widetilde{T})^\ast \rho(\widetilde{CP})^{-1}
    &= \rho(\widetilde{T})^{-1},
\end{align}
are satisfied for matter wavefunctions on $T^2$ as well as $T^2/\mathbb{Z}_2$, where the latter $\mathbb{Z}_2$ case can be explicitly checked via the representations below. Thus, we constructed the outer automorphism $u_{\rm CP}: G_{\rm CP} \rightarrow {\rm Aut}(G_{\rm modular}$).  Here, $G_{\rm modular}$ denotes the finite metaplectic modular group that the wavefunctions enjoy. 
In the semi-direct product group, the generators satisfy
\begin{align}
    \widetilde{CP} \widetilde{S} \widetilde{CP}^{-1} = \widetilde{S}^{-1}, \nonumber \\
    \widetilde{CP} \widetilde{T} \widetilde{CP}^{-1} = \widetilde{T}^{-1},
\end{align}
where the generators are interpreted as elements of $G_{\rm modular} \rtimes G_{\rm CP}$.

\medskip

On the $T^2/\mathbb{Z}_2$ orbifold, the explicit forms of the modular (flavor) transformations are summarized as follows:
\begin{itemize}
\item $M=2$

Since there is no $\mathbb{Z}_2$-odd mode, the modular transformations of $\mathbb{Z}_2$-even mode are given by
\begin{align}
    \rho(\widetilde{S}_{\rm even}) = 
    -\frac{e^{-\frac{i\pi}{4}}}{\sqrt{2}}
    \begin{pmatrix}
        -1 & -1\\
        -1 & 1 \\
    \end{pmatrix}
    ,\qquad
    \rho(\widetilde{T}_{\rm even}) =     
    \begin{pmatrix}
        1 & 0\\
        0 & -i \\
    \end{pmatrix}
    ,
\end{align}
which are regarded as the representation matrices of $\Tilde{\Gamma}_4$. 

\item $M=3$

There are two $\mathbb{Z}_2$-even modes and single $\mathbb{Z}_2$-odd mode. 
Their modular transformations are given by
\begin{align}
    \rho(\widetilde{S}_{\rm even}) &= 
    -\frac{-i}{\sqrt{3}}
    \begin{pmatrix}
        1 & \sqrt{2}\\
        \sqrt{2} & -1 \\
    \end{pmatrix}
    ,\qquad
    \rho(\widetilde{T}_{\rm even}) =     
    \begin{pmatrix}
        1 & 0\\
        0 & -(-1)^{1/3} \\
    \end{pmatrix}
    ,
    \nonumber\\
    \rho(\widetilde{S}_{\rm odd}) &=1,
    \qquad
    \rho(\widetilde{T}_{\rm odd}) = -(-1)^{1/3},    
\end{align}
which are regarded as the representation matrices of $\Tilde{\Gamma}_{12}$.

\item $M=4$

There are two $\mathbb{Z}_2$-even modes and single $\mathbb{Z}_2$-odd mode. 
Their modular transformations are given by
\begin{align}
    \rho(\widetilde{S}_{\rm even}) &= 
    -\frac{1}{2}
    \begin{pmatrix}
        (-1)^{1/4} & 1+i & (-1)^{1/4}\\
         1+i & 0 & -1-i\\
        (-1)^{1/4} & -1-i & (-1)^{1/4}\\
    \end{pmatrix}
    ,\qquad
    \rho(\widetilde{T}_{\rm even}) =     
    \begin{pmatrix}
        1 & 0 & 0\\
        0 & -(-1)^{1/4} & 0 \\
        0 & 0 & -1\\
    \end{pmatrix}
    ,
    \nonumber\\
    \rho(\widetilde{S}_{\rm odd}) &=(-1)^{3/4},
    \qquad
    \rho(\widetilde{T}_{\rm odd}) = -(-1)^{1/4},    
    \label{eq:reptildegamma8}
\end{align}
which are regarded as the representation matrices of $\Tilde{\Gamma}_8$.

\item $M=5$

There are two $\mathbb{Z}_2$-even modes and single $\mathbb{Z}_2$-odd mode. 
Their modular transformations are given by
\begin{align}
    \rho(\widetilde{S}_{\rm even}) &= 
    -\frac{1}{\sqrt{5}}
    \begin{pmatrix}
        1 & \sqrt{2} & \sqrt{2}\\
        \sqrt{2} & \frac{-1 + \sqrt{5}}{2} & \frac{-1 - \sqrt{5}}{2}\\
       \sqrt{2} & \frac{-1 - \sqrt{5}}{2} & \frac{-1 + \sqrt{5}}{2}\\
    \end{pmatrix}
    ,\qquad
    \rho(\widetilde{T}_{\rm even}) =     
    \begin{pmatrix}
        1 & 0 & 0\\
        0 & -(-1)^{1/5} & 0 \\
        0 & 0 & (-1)^{4/5}\\
    \end{pmatrix}
    ,
    \nonumber\\
    \rho(\widetilde{S}_{\rm odd}) &=
        \begin{pmatrix}
        -1 & 0\\
        0 & -1 \\
    \end{pmatrix}
    ,
    \qquad
    \rho(\widetilde{T}_{\rm odd}) = 
        \begin{pmatrix}
        -(-1)^{1/5} & 0\\
        0 & (-1)^{4/5} \\
    \end{pmatrix}
    ,
\end{align}
which will be regarded as the representation matrices of $\Tilde{\Gamma}_{20}$.

\item $M=6$

There are four $\mathbb{Z}_2$-even modes and two $\mathbb{Z}_2$-odd modes. 
Their modular transformations are given by
\begin{align}
    \rho(\widetilde{S}_{\rm even}) &= 
    \frac{1}{\sqrt{6}}
    \begin{pmatrix}
        \frac{1-i}{\sqrt{2}} & 1-i & 1-i &  \frac{1-i}{\sqrt{2}}\\
        1-i & \frac{1-i}{\sqrt{2}} & (-1)^{3/4} & -1+i\\
        1-i & (-1)^{3/4} & (-1)^{3/4} & 1-i\\
        \frac{1-i}{\sqrt{2}} & -1+i & 1-i &  (-1)^{3/4}\\
    \end{pmatrix}
    ,
    \nonumber\\
    \rho(\widetilde{T}_{\rm even}) &=     
    \begin{pmatrix}
        1 & 0 & 0 & 0\\
        0 & -(-1)^{1/6} & 0 \\
        0 & 0 & (-1)^{2/3} & 0\\
        0 & 0 & 0 & i\\
    \end{pmatrix}
    ,
    \nonumber\\
    \rho(\widetilde{S}_{\rm odd}) &=
    \frac{1}{2}
        \begin{pmatrix}
        1+i & 1+i\\
        1+i & -1-i \\
    \end{pmatrix}
    ,
    \qquad
    \rho(\widetilde{T}_{\rm odd}) = 
        \begin{pmatrix}
        -(-1)^{1/6} & 0\\
        0 & (-1)^{2/3} \\
    \end{pmatrix}
    ,
\end{align}
which are regarded as the representation matrices of $\Tilde{\Gamma}_{12}$.

\end{itemize}

\subsection{Eclectic flavor symmetry}
\label{sec:10D}

In this section, we analyze the Type IIB magnetized D-brane models in Sec. \ref{sec:generation2}, 
where the three generation models with Pati-Salam gauge symmetry are realized in the visible sector. 
Since the generation number of quarks, leptons, and Higgs doublets is determined by magnetic fluxes on the first torus $(T^2)_1$, 
\begin{align}
    I_{ab} = I_{ab}^1I_{ab}^2I_{ab}^3 =g(-1)(-1)=g,\qquad
    I_{ca} = I_{ca}^1I_{ca}^2I_{ca}^3 = (-g)(-1)1=g,
\end{align}
the flavor structure is derived from the first torus. 
Indeed, the wavefunctions of matter fields on second and third tori are just the constant. 
Note that $I_{bc}^3=0$ indicates that Higgs doublets are vector-like particles 
due to $I_{bc}^3=0$. Thus, the index of Higgs doublets is calculated by using Eq.(\ref{eq:index_Higgs}) with $s_1<0, s_2>0$ under the assumption $g>0$:
\begin{align}
    I_{bc} = -g+1.
\end{align}
From the results of Sec. \ref{sec:generation2}, it turned out that the string landscape leads to a few number of generation of quarks and lepton. 
In the following analysis, we thus focus on matter wavefunctions on $(T^2)_1$ with vanishing Wilson lines whose explicit forms are given in Eq. (\ref{eq:wave_evenodd}) with
\begin{align}
    \psi^{\talpha, |M|}(z,\tau) = \left( \frac{|M|}{{\cal A}^2}\right)^{1/4} e^{i\pi |M| z\frac{{\rm Im}z}{{\rm Im}\tau}}
      \vartheta \begin{bmatrix}
    \frac{\talpha}{|M|} \\
    0 \\
  \end{bmatrix}
  (|M|z,|M|\tau), 
  \label{eq:wave_meta}
\end{align}
with $M= I_{ab}, I_{bc}, I_{ca}$. 
Note that the orbifold projections split the wavefunctions to $\mathbb{Z}_2$-even and $\mathbb{Z}_2$-odd modes. 
For illustrative purposes, we focus on traditional flavor symmetries on three $\mathbb{Z}_2$-even modes with $g=4$.\footnote{See for the unification of traditional flavor and modular symmetries on $T^2$ with magnetic fluxes \cite{Ohki:2020bpo}.} 

\medskip

\begin{itemize}
    \item $g=4$

The traditional flavor symmetries are described by 
$G_{\rm flavor}:=\mathbb{Z}_4\times \mathbb{Z}_2^{(P)} \times \mathbb{Z}_2^{(C)} \times \mathbb{Z}_2^{(Z)}$ 
whose generators $\{Z', P, C, Z\}$ are of the form \cite{Abe:2009vi}:
\begin{align}
    \rho(Z'_{\rm even}) = i\mathbb{I}_3,
    \quad
    \rho(P_{\rm even}) = \mathbb{I}_3,
    \quad
   \rho(C_{\rm even}) =
    \begin{pmatrix}
        0 & 0 & 1\\
        0 & 1 & 0\\
        1 & 0 & 0\\
    \end{pmatrix}
    ,
    \quad
    \rho(Z_{\rm even}) =
    \begin{pmatrix}
        1 & 0 & 0\\
        0 & -1 & 0\\
        0 & 0 & 1\\
    \end{pmatrix}
    ,
\end{align}
for $\mathbb{Z}_2$-even mode 
and
\begin{align}
    \rho(Z'_{\rm odd}) = i,
    \qquad
    \rho(P_{\rm odd}) = -1
    \qquad
    \rho(C_{\rm odd}) =-1
    ,
    \qquad
    \rho(Z_{\rm odd}) =-1
    ,
\end{align}
for $\mathbb{Z}_2$-odd mode. 
Note that such flavor symmetries do not change Yukawa couplings and only act on 
three generations of quarks and leptons.

\medskip

In this case, the wavefunctions on $T^2/Z_2$ enjoy the modular flavor group $\widetilde{\Gamma}_8$ and the explicit representations are given in Eq. (\ref{eq:reptildegamma8}).
In particular, the flavor generators do not commute with those of modular flavor symmetries $G_{\rm modular}=\widetilde{\Gamma}_8$. 
Indeed, we find that
\begin{align}
    \widetilde{S}_{\rm even}  C_{\rm even} \widetilde{S}_{\rm even} ^{-1} &= Z_{\rm even},
    \quad
    \widetilde{S}_{\rm even}  Z_{\rm even} \widetilde{S}_{\rm even} ^{-1} = C_{\rm even},
    \nonumber\\
    \widetilde{T}_{\rm even}  C_{\rm even} \widetilde{T}_{\rm even} ^{-1} &= C_{\rm even}Z_{\rm even}(Z'_{\rm even})^2,
    \quad
    \widetilde{T}_{\rm even}  Z_{\rm even} \widetilde{T}_{\rm even}^{-1} = Z_{\rm even},
\end{align}
and the other generators of $G_{\rm flavor}$ commute with $G_{\rm modular}$. 
It means that the modular transformation is regarded as an automorphism of the traditional flavor group.

\medskip

Furthermore, we can construct the outer automorphism $u_{\rm CP}: G_{\rm CP} \rightarrow {\rm Aut} (G_{\rm flavor} \rtimes G_{\rm modular})$. 
Indeed, the following relations can be verified in the semi-direct product group:
\begin{align}
            \widetilde{CP}Z'_{\rm even} \widetilde{CP}^{-1} &= (Z'_{\rm even})^{-1}, \quad \widetilde{CP} Z'_{\rm odd} \widetilde{CP}^{-1} = (Z'_{\rm odd})^{-1}, \nonumber \\
    \widetilde{CP} \widetilde{S} \widetilde{CP}^{-1} &= \widetilde{S}^{-1}, \quad \widetilde{CP} \widetilde{T} \widetilde{CP}^{-1} = \widetilde{T}^{-1}.
\end{align}
Recalling that the modular flavor and CP symmetries are treated in a uniform manner, the traditional flavor, modular flavor and CP symmetries are described by
\begin{align}
    (G_{\rm flavor} \rtimes G_{\rm modular})\rtimes G_{\rm CP},
\end{align}
as discussed in heterotic orbifold models. 
From the analysis of Sec. \ref{sec:2}, the moduli fields can be stabilized in flux compactifications, in particular, $\mathbb{Z}_3$ fixed point. 
It leads to $\mathbb{Z}_3$ modular symmetry generated by $\{1,ST,(ST)^2\}$. It turned out that such a $\mathbb{Z}_3$ symmetry still enhances the flavor symmetry due to the relation:
\begin{align}
    (\widetilde{S}\widetilde{T})C_{\rm even}(\widetilde{S}\widetilde{T})^{-1}
    = C_{\rm even}Z_{\rm even}(Z'_{\rm even})^2,
    \qquad
    (\widetilde{S}\widetilde{T})Z_{\rm even}(\widetilde{S}\widetilde{T})^{-1}
    = Z_{\rm even}.
\end{align}
Thus, the discrete non-abelian symmetry $(G_{\rm flavor} \rtimes \mathbb{Z}_3)\rtimes G_{\rm CP}$ remains in the low-energy action. 
So far, we have focused on specific magnetized D-brane models with stabilized moduli, 
but it is quite interesting to explore other flavor models, which left for future work.

\end{itemize}

\medskip

\section{Conclusions}
\label{sec:con}

In this paper, we have examined the vacuum structure of Type IIB flux vacua with SM spectra. 
The background fluxes play an important role in stabilizing moduli fields and determining the generation 
number of chiral zero-modes. 
Since the background fluxes are constrained by the tadpole cancellation conditions, 
the moduli distribution and the generation number are mutually related with each other. 
By studying the $T^6/(\mathbb{Z}_2\times \mathbb{Z}_2')$ orientifolds with magnetized 
D-brane models in Secs. \ref{sec:generation1} and \ref{sec:generation2}, 
it is found that the string landscape leads to the small generation number of quarks and leptons. 
Furthermore, the moduli values are peaked at $\mathbb{Z}_3$ fixed point in the complex structure 
moduli space. 
It motivates us to study whether such a discrete symmetry is related to the flavor and/or CP symmetries in the low-energy effective action. 

\medskip

To investigate the relation between the modular symmetry of the torus and the flavor symmetries of 
quarks and leptons, we have focused on the concrete magnetized D-brane model of Sec. \ref{sec:generation2}. 
Since the wavefunctions of chiral zero-modes and the corresponding Yukawa couplings are 
written by Jacobi theta function with the modular weight $1/2$, 
they are described in the framework of metaplectic modular flavor symmetry. 
Note that the flavor structure of quarks and leptons is originated from one of tori. 
We found that the modular transformations of both the $\mathbb{Z}_2$-even and -odd modes are 
described by $\Tilde{\Gamma}_{2|M|}$ and $\Tilde{\Gamma}_{4|M|}$ for the magnetic flux with even $|M|$ and odd $|M|$, respectively, 
Furthermore, the CP symmetry can be regarded as the outer automorphism of the metaplectic modular group. 
For illustrative purposes, we focus on the $M=4$ case, 
where three $\mathbb{Z}_2$-even modes transform under a certain traditional flavor symmetry. 
We found that the traditional flavor, modular flavor and CP symmetries in Type IIB chiral flux vacua are uniformly described in the context of 
eclectic flavor symmetry: $(G_{\rm flavor} \rtimes G_{\rm modular})\rtimes G_{\rm CP}$ as discussed in 
the heterotic orbifolds \cite{Nilles:2020nnc,Nilles:2020kgo}. 
It would be interesting to explore the realization of eclectic flavor symmetry on other corners of string models. 

\medskip

Furthermore, we have stabilized the moduli fields in the framework of flux compactifications. 
Although the moduli vacuum expectation values are distributed around $\mathbb{Z}_3$ fixed point\footnote{If 
we consider the stabilization of K\"ahler moduli by non-perturbative effects, the $\mathbb{Z}_3$ symmetry will be 
slightly broken as analyzed in Ref. \cite{Ishiguro:2022pde}.}, 
a part of eclectic flavor symmetry $(G_{\rm flavor} \rtimes G_{\rm modular})\rtimes G_{\rm CP}$ still 
remains in the low-energy effective action. 
Since the coefficient of 4D higher-dimensional operators will be described by the product of modular forms with 
half-integer modular weights, the eclectic flavor symmetry would control the flavor structure 
of higher-dimensional operators. 
We leave a pursue of these interesting topics for future work.

\acknowledgments

This work was supported by JSPS KAKENHI Grant Numbers JP20K14477 (Hajime O.), JP22J12877 (K.I) and JP23H04512 (Hajime O.).
The work of Hiroshi O. is supported by the Junior Research Group (JRG) Program at the Asia-Pacific Center for Theoretical
Physics (APCTP) through the Science and Technology Promotion Fund and Lottery Fund of the Korean Government and was supported by the Korean Local Governments-Gyeongsangbuk-do Province and Pohang City. 
Hiroshi O. is sincerely grateful for all the KIAS members.

\appendix

\section{Modular symmetries in flux compactifications}
\label{app}

In this Appendix, we confirm the modular invariance of the effective action 
in the context of flux compactifications. 
The relevant superpotential and the K\"ahler potential in the isotropic regime (\ref{eq:isotropic}) are given by
\begin{align}
    K &= -3\ln (i(\Bar{\tau}-\tau))-\ln (i(\Bar{S}-S)) -2\ln {\cal V}(T,\Bar{T}),
    \nonumber\\
    W &= a^0 \tau^3- 3a\tau^2- 3b \tau-b_0 -S\left(c^0\tau^3-3c\tau^2-3d\tau-d_0 \right).  
\end{align}

The effective action is invariant under the following three classes of transformations which are enumerated as follows. 

\begin{enumerate}
    \item $SL(2,\mathbb{Z})_\tau$ modular symmetry

Following Ref. \cite{DeWolfe:2004ns}, let us discuss the $SL(2,\mathbb{Z})_{\tau_i}$ modular symmetry of the effective action. 
The period vector $\Pi(\tau_i) :=\{X^I, F_I\}$ transforms under the transformation of 
moduli fields $\tau_i \rightarrow \tau_i^\prime$:
\begin{align}
    \Pi(\tau_i) \rightarrow \Lambda (\tau_i)\,U \cdot \Pi(\tau_i),
\end{align}
where $\Lambda$ denotes a moduli-dependent function and $U$ is a $Sp(8,\mathbb{Z})$ symplectic matrix. 
The explicit form of the symplectic matrix is given in Eq. (\ref{eq:S1T1}). 
Such a transformation of the period vector is nothing but the K\"ahler transformation:
\begin{align}
    K \rightarrow K - \ln \Lambda - \ln \Bar{\Lambda},\qquad
    W \rightarrow \Lambda W,
\end{align}
under which the fluxes $f= \{ \int_{A^I} F_3, \int_{B_I} F_3\}$ and $H= \{ \int_{A^I} H_3, \int_{B_I} H_3\}$ transform as
\begin{align}
    f \rightarrow f \cdot U^{-1},\qquad h \rightarrow h \cdot U^{-1}.
\end{align}
The $S$- and $T$-transformations of flux quanta are of the explicit form \cite{Betzler:2019kon,Ishiguro:2020tmo}:
\begin{itemize}
\item $S: \tau \rightarrow -1/\tau$
\begin{align}
a^0\rightarrow b_0,\quad a\rightarrow b,\quad b\rightarrow -a,\quad b_0 \rightarrow -a^0,
\\ \notag
c^0\rightarrow d_0,\quad c\rightarrow d,\quad d\rightarrow -c,\quad d_0 \rightarrow -c^0.
\end{align}
\item $T: \tau \rightarrow \tau+1$
\begin{align}
&a^0\rightarrow a^0,\quad a\rightarrow a + a^0,\quad b\rightarrow b-2a-a^0,\quad b_0 \rightarrow b_0-3b +3a +a^0,
\\ \notag
&c^0\rightarrow c^0,\quad c\rightarrow c + c^0,\quad d\rightarrow d-2c-c^0,\quad d_0 \rightarrow d_0-3d+3c+c^0.
\end{align}
\end{itemize}

    \item $SL(2,\mathbb{Z})_S$ modular symmetry

The axion-dilaton transforms under the $SL(2,\mathbb{Z})_S$ modular symmetry:
\begin{align}
    S \rightarrow S^\prime =\frac{pS+q}{sS +t}=R_S(\tau_i),
\end{align}
with $p t -qs=1$. 
Then, the effective action enjoys the modular symmetry due to the transformation of the RR and NS fluxes:
\begin{align}
    \begin{pmatrix}
    F\\
    H
     \end{pmatrix}
     \rightarrow
    \begin{pmatrix}
    p & q\\
    s & t
    \end{pmatrix}
     \begin{pmatrix}
    F\\
    H
     \end{pmatrix}   
.
\end{align}
It corresponds to the following transformations of the flux quanta under the $S$- and $T$-transformations of $SL(2, \mathbb{Z})_S$:
\begin{itemize}
    \item $S: S \rightarrow -1/S$ 
    \begin{align}
&a^0\rightarrow -c^0,\quad a\rightarrow -c,\quad b\rightarrow -d,\quad b_0 \rightarrow -d_0,
\\ \notag
&c^0\rightarrow a^0,\quad c\rightarrow a,\quad d\rightarrow b,\quad d_0 \rightarrow b_0.
\end{align}
\item $T: S \rightarrow S+1$
\begin{align}
&a^0\rightarrow a^0+c^0,\quad a\rightarrow a+c,\quad b\rightarrow b+d,\quad b_0 \rightarrow b_0+d_0,
\\ \notag
&c^0\rightarrow c^0,\quad c\rightarrow c,\quad d\rightarrow d,\quad d_0 \rightarrow d_0.
    \end{align}
\end{itemize}

    \item Flipping the overall sign of flux quanta

When all the flux quanta flip the sign:
\begin{align}
(a^0, a, b, b_0, c^0, c, d, d_0)
\rightarrow
-(a^0, a, b, b_0, c^0, c, d, d_0),
\label{eq:third}
\end{align}
they cause the sign flipping of the superpotential $W\rightarrow -W$, 
but the effective action is still invariant up to the the K\"ahler transformation. 
Such a sign flipping reduces $SL(2, \mathbb{Z})_{\tau, S}$ to $PSL(2, \mathbb{Z})_{\tau, S}$. 

\end{enumerate}

These three transformations play the role in counting the finite physically-distinct vacua as analyzed in Section \ref{sec:2}. 
Note that the flux-induced D3-brane charge (\ref{eq:nD3}) is also invariant under these transformations.

\bibliography{referencesv2}{}

\providecommand{\href}[2]{#2}\begingroup\raggedright\begin{thebibliography}{10}

\bibitem{Ashok:2003gk}
S.~Ashok and M.R.~Douglas, \emph{{Counting flux vacua}},
  \href{https://doi.org/10.1088/1126-6708/2004/01/060}{\emph{JHEP} {\bfseries
  01} (2004) 060} [\href{https://arxiv.org/abs/hep-th/0307049}{{\ttfamily
  hep-th/0307049}}].

\bibitem{Douglas:2003um}
M.R.~Douglas, \emph{{The Statistics of string / M theory vacua}},
  \href{https://doi.org/10.1088/1126-6708/2003/05/046}{\emph{JHEP} {\bfseries
  05} (2003) 046} [\href{https://arxiv.org/abs/hep-th/0303194}{{\ttfamily
  hep-th/0303194}}].

\bibitem{Denef:2004ze}
F.~Denef and M.R.~Douglas, \emph{{Distributions of flux vacua}},
  \href{https://doi.org/10.1088/1126-6708/2004/05/072}{\emph{JHEP} {\bfseries
  05} (2004) 072} [\href{https://arxiv.org/abs/hep-th/0404116}{{\ttfamily
  hep-th/0404116}}].

\bibitem{Vafa:2005ui}
C.~Vafa, \emph{{The String landscape and the swampland}},
  \href{https://arxiv.org/abs/hep-th/0509212}{{\ttfamily hep-th/0509212}}.

\bibitem{ArkaniHamed:2006dz}
N.~Arkani-Hamed, L.~Motl, A.~Nicolis and C.~Vafa, \emph{{The String landscape,
  black holes and gravity as the weakest force}},
  \href{https://doi.org/10.1088/1126-6708/2007/06/060}{\emph{JHEP} {\bfseries
  06} (2007) 060} [\href{https://arxiv.org/abs/hep-th/0601001}{{\ttfamily
  hep-th/0601001}}].

\bibitem{Ooguri:2006in}
H.~Ooguri and C.~Vafa, \emph{{On the Geometry of the String Landscape and the
  Swampland}},
  \href{https://doi.org/10.1016/j.nuclphysb.2006.10.033}{\emph{Nucl. Phys. B}
  {\bfseries 766} (2007) 21}
  [\href{https://arxiv.org/abs/hep-th/0605264}{{\ttfamily hep-th/0605264}}].

\bibitem{Palti:2019pca}
E.~Palti, \emph{{The Swampland: Introduction and Review}},
  \href{https://doi.org/10.1002/prop.201900037}{\emph{Fortsch. Phys.}
  {\bfseries 67} (2019) 1900037}
  [\href{https://arxiv.org/abs/1903.06239}{{\ttfamily 1903.06239}}].

\bibitem{DeWolfe:2004ns}
O.~DeWolfe, A.~Giryavets, S.~Kachru and W.~Taylor, \emph{{Enumerating flux
  vacua with enhanced symmetries}},
  \href{https://doi.org/10.1088/1126-6708/2005/02/037}{\emph{JHEP} {\bfseries
  02} (2005) 037} [\href{https://arxiv.org/abs/hep-th/0411061}{{\ttfamily
  hep-th/0411061}}].

\bibitem{Ishiguro:2020tmo}
K.~Ishiguro, T.~Kobayashi and H.~Otsuka, \emph{{Landscape of Modular Symmetric
  Flavor Models}}, \href{https://doi.org/10.1007/JHEP03(2021)161}{\emph{JHEP}
  {\bfseries 03} (2021) 161}
  [\href{https://arxiv.org/abs/2011.09154}{{\ttfamily 2011.09154}}].

\bibitem{Cremades:2004wa}
D.~Cremades, L.E.~Ibanez and F.~Marchesano, \emph{{Computing Yukawa couplings
  from magnetized extra dimensions}},
  \href{https://doi.org/10.1088/1126-6708/2004/05/079}{\emph{JHEP} {\bfseries
  05} (2004) 079} [\href{https://arxiv.org/abs/hep-th/0404229}{{\ttfamily
  hep-th/0404229}}].

\bibitem{Nilles:2020nnc}
H.P.~Nilles, S.~Ramos-S\'anchez and P.K.S.~Vaudrevange, \emph{{Eclectic Flavor
  Groups}}, \href{https://doi.org/10.1007/JHEP02(2020)045}{\emph{JHEP}
  {\bfseries 02} (2020) 045}
  [\href{https://arxiv.org/abs/2001.01736}{{\ttfamily 2001.01736}}].

\bibitem{Nilles:2020kgo}
H.P.~Nilles, S.~Ramos-Sanchez and P.K.S.~Vaudrevange, \emph{{Lessons from
  eclectic flavor symmetries}},
  \href{https://doi.org/10.1016/j.nuclphysb.2020.115098}{\emph{Nucl. Phys. B}
  {\bfseries 957} (2020) 115098}
  [\href{https://arxiv.org/abs/2004.05200}{{\ttfamily 2004.05200}}].

\bibitem{Baur:2020jwc}
A.~Baur, M.~Kade, H.P.~Nilles, S.~Ramos-Sanchez and P.K.S.~Vaudrevange,
  \emph{{The eclectic flavor symmetry of the $\boldsymbol{\mathbb{Z}_2}$
  orbifold}}, \href{https://doi.org/10.1007/JHEP02(2021)018}{\emph{JHEP}
  {\bfseries 02} (2021) 018}
  [\href{https://arxiv.org/abs/2008.07534}{{\ttfamily 2008.07534}}].

\bibitem{Baur:2022hma}
A.~Baur, H.P.~Nilles, S.~Ramos-Sanchez, A.~Trautner and P.K.S.~Vaudrevange,
  \emph{{The first string-derived eclectic flavor model with realistic
  phenomenology}}, \href{https://doi.org/10.1007/JHEP09(2022)224}{\emph{JHEP}
  {\bfseries 09} (2022) 224}
  [\href{https://arxiv.org/abs/2207.10677}{{\ttfamily 2207.10677}}].

\bibitem{Nilles:2020gvu}
H.P.~Nilles, S.~Ramos\textendash{}S\'anchez and P.K.S.~Vaudrevange,
  \emph{{Eclectic flavor scheme from ten-dimensional string theory - II
  detailed technical analysis}},
  \href{https://doi.org/10.1016/j.nuclphysb.2021.115367}{\emph{Nucl. Phys. B}
  {\bfseries 966} (2021) 115367}
  [\href{https://arxiv.org/abs/2010.13798}{{\ttfamily 2010.13798}}].

\bibitem{DNilles:2020gvu}
G.-J.~Ding, S.F.~King, C.-C.~Li, X.-G.~Liu and J.-N.~Lu, \emph{{Neutrino mass
  and mixing models with eclectic flavor symmetry \ensuremath{\Delta}(27)
  \ensuremath{\rtimes} T'}},
  \href{https://doi.org/10.1007/JHEP05(2023)144}{\emph{JHEP} {\bfseries 05}
  (2023) 144} [\href{https://arxiv.org/abs/2303.02071}{{\ttfamily
  2303.02071}}].

\bibitem{Blumenhagen:2006ci}
R.~Blumenhagen, B.~Kors, D.~Lust and S.~Stieberger, \emph{{Four-dimensional
  String Compactifications with D-Branes, Orientifolds and Fluxes}},
  \href{https://doi.org/10.1016/j.physrep.2007.04.003}{\emph{Phys. Rept.}
  {\bfseries 445} (2007) 1}
  [\href{https://arxiv.org/abs/hep-th/0610327}{{\ttfamily hep-th/0610327}}].

\bibitem{Ishiguro:2021ccl}
K.~Ishiguro, T.~Kobayashi and H.~Otsuka, \emph{{Symplectic modular symmetry in
  heterotic string vacua: flavor, CP, and R-symmetries}},
  \href{https://doi.org/10.1007/JHEP01(2022)020}{\emph{JHEP} {\bfseries 01}
  (2022) 020} [\href{https://arxiv.org/abs/2107.00487}{{\ttfamily
  2107.00487}}].

\bibitem{Gukov:1999ya}
S.~Gukov, C.~Vafa and E.~Witten, \emph{{CFT's from Calabi-Yau four folds}},
  \href{https://doi.org/10.1016/S0550-3213(00)00373-4}{\emph{Nucl. Phys. B}
  {\bfseries 584} (2000) 69}
  [\href{https://arxiv.org/abs/hep-th/9906070}{{\ttfamily hep-th/9906070}}].

\bibitem{Ishiguro:2020nuf}
K.~Ishiguro, T.~Kobayashi and H.~Otsuka, \emph{{Spontaneous CP violation and
  symplectic modular symmetry in Calabi-Yau compactifications}},
  \href{https://doi.org/10.1016/j.nuclphysb.2021.115598}{\emph{Nucl. Phys. B}
  {\bfseries 973} (2021) 115598}
  [\href{https://arxiv.org/abs/2010.10782}{{\ttfamily 2010.10782}}].

\bibitem{Ishiguro:2022pde}
K.~Ishiguro, H.~Okada and H.~Otsuka, \emph{{Residual flavor symmetry breaking
  in the landscape of modular flavor models}},
  \href{https://doi.org/10.1007/JHEP09(2022)072}{\emph{JHEP} {\bfseries 09}
  (2022) 072} [\href{https://arxiv.org/abs/2206.04313}{{\ttfamily
  2206.04313}}].

\bibitem{Kobayashi:2020hoc}
T.~Kobayashi and H.~Otsuka, \emph{{Classification of discrete modular
  symmetries in Type IIB flux vacua}},
  \href{https://doi.org/10.1103/PhysRevD.101.106017}{\emph{Phys. Rev. D}
  {\bfseries 101} (2020) 106017}
  [\href{https://arxiv.org/abs/2001.07972}{{\ttfamily 2001.07972}}].

\bibitem{Candelas:1997eh}
P.~Candelas, E.~Perevalov and G.~Rajesh, \emph{{Toric geometry and enhanced
  gauge symmetry of F theory / heterotic vacua}},
  \href{https://doi.org/10.1016/S0550-3213(97)00563-4}{\emph{Nucl. Phys. B}
  {\bfseries 507} (1997) 445}
  [\href{https://arxiv.org/abs/hep-th/9704097}{{\ttfamily hep-th/9704097}}].

\bibitem{Taylor:2015xtz}
W.~Taylor and Y.-N.~Wang, \emph{{The F-theory geometry with most flux vacua}},
  \href{https://doi.org/10.1007/JHEP12(2015)164}{\emph{JHEP} {\bfseries 12}
  (2015) 164} [\href{https://arxiv.org/abs/1511.03209}{{\ttfamily
  1511.03209}}].

\bibitem{Marchesano:2013ega}
F.~Marchesano, D.~Regalado and L.~Vazquez-Mercado, \emph{{Discrete flavor
  symmetries in D-brane models}},
  \href{https://doi.org/10.1007/JHEP09(2013)028}{\emph{JHEP} {\bfseries 09}
  (2013) 028} [\href{https://arxiv.org/abs/1306.1284}{{\ttfamily 1306.1284}}].

\bibitem{Witten:1982fp}
E.~Witten, \emph{{An SU(2) Anomaly}},
  \href{https://doi.org/10.1016/0370-2693(82)90728-6}{\emph{Phys. Lett. B}
  {\bfseries 117} (1982) 324}.

\bibitem{Uranga:2000xp}
A.M.~Uranga, \emph{{D-brane probes, RR tadpole cancellation and K-theory
  charge}}, \href{https://doi.org/10.1016/S0550-3213(00)00787-2}{\emph{Nucl.
  Phys. B} {\bfseries 598} (2001) 225}
  [\href{https://arxiv.org/abs/hep-th/0011048}{{\ttfamily hep-th/0011048}}].

\bibitem{Marchesano:2004xz}
F.~Marchesano and G.~Shiu, \emph{{Building MSSM flux vacua}},
  \href{https://doi.org/10.1088/1126-6708/2004/11/041}{\emph{JHEP} {\bfseries
  11} (2004) 041} [\href{https://arxiv.org/abs/hep-th/0409132}{{\ttfamily
  hep-th/0409132}}].

\bibitem{Kumar:2005hf}
J.~Kumar and J.D.~Wells, \emph{{Surveying standard model flux vacua on T**6 /
  Z(2) x Z(2)}},
  \href{https://doi.org/10.1088/1126-6708/2005/09/067}{\emph{JHEP} {\bfseries
  09} (2005) 067} [\href{https://arxiv.org/abs/hep-th/0506252}{{\ttfamily
  hep-th/0506252}}].

\bibitem{Kachru:2002gs}
S.~Kachru, J.~Pearson and H.L.~Verlinde, \emph{{Brane / flux annihilation and
  the string dual of a nonsupersymmetric field theory}},
  \href{https://doi.org/10.1088/1126-6708/2002/06/021}{\emph{JHEP} {\bfseries
  06} (2002) 021} [\href{https://arxiv.org/abs/hep-th/0112197}{{\ttfamily
  hep-th/0112197}}].

\bibitem{Cvetic:2005bn}
M.~Cvetic, T.~Li and T.~Liu, \emph{{Standard-like models as type IIB flux
  vacua}}, \href{https://doi.org/10.1103/PhysRevD.71.106008}{\emph{Phys. Rev.
  D} {\bfseries 71} (2005) 106008}
  [\href{https://arxiv.org/abs/hep-th/0501041}{{\ttfamily hep-th/0501041}}].

\bibitem{Blumenhagen:2000ea}
R.~Blumenhagen, B.~Kors and D.~Lust, \emph{{Type I strings with F flux and B
  flux}}, \href{https://doi.org/10.1088/1126-6708/2001/02/030}{\emph{JHEP}
  {\bfseries 02} (2001) 030}
  [\href{https://arxiv.org/abs/hep-th/0012156}{{\ttfamily hep-th/0012156}}].

\bibitem{Cascales:2003zp}
J.F.G.~Cascales and A.M.~Uranga, \emph{{Chiral 4d string vacua with D branes
  and NSNS and RR fluxes}},
  \href{https://doi.org/10.1088/1126-6708/2003/05/011}{\emph{JHEP} {\bfseries
  05} (2003) 011} [\href{https://arxiv.org/abs/hep-th/0303024}{{\ttfamily
  hep-th/0303024}}].

\bibitem{Cvetic:2001nr}
M.~Cvetic, G.~Shiu and A.M.~Uranga, \emph{{Chiral four-dimensional N=1
  supersymmetric type 2A orientifolds from intersecting D6 branes}},
  \href{https://doi.org/10.1016/S0550-3213(01)00427-8}{\emph{Nucl. Phys. B}
  {\bfseries 615} (2001) 3}
  [\href{https://arxiv.org/abs/hep-th/0107166}{{\ttfamily hep-th/0107166}}].

\bibitem{Chen:2007px}
C.-M.~Chen, T.~Li, V.E.~Mayes and D.V.~Nanopoulos, \emph{{A Realistic world
  from intersecting D6-branes}},
  \href{https://doi.org/10.1016/j.physletb.2008.06.024}{\emph{Phys. Lett. B}
  {\bfseries 665} (2008) 267}
  [\href{https://arxiv.org/abs/hep-th/0703280}{{\ttfamily hep-th/0703280}}].

\bibitem{Liu:2020msy}
X.-G.~Liu, C.-Y.~Yao, B.-Y.~Qu and G.-J.~Ding, \emph{{Half-integral weight
  modular forms and application to neutrino mass models}},
  \href{https://doi.org/10.1103/PhysRevD.102.115035}{\emph{Phys. Rev. D}
  {\bfseries 102} (2020) 115035}
  [\href{https://arxiv.org/abs/2007.13706}{{\ttfamily 2007.13706}}].

\bibitem{GAP4}
The GAP~Group, \emph{{GAP -- Groups, Algorithms, and Programming, Version
  4.12.2}}, 2022.

\bibitem{Kobayashi:2018rad}
T.~Kobayashi, S.~Nagamoto, S.~Takada, S.~Tamba and T.H.~Tatsuishi,
  \emph{{Modular symmetry and non-Abelian discrete flavor symmetries in string
  compactification}},
  \href{https://doi.org/10.1103/PhysRevD.97.116002}{\emph{Phys. Rev. D}
  {\bfseries 97} (2018) 116002}
  [\href{https://arxiv.org/abs/1804.06644}{{\ttfamily 1804.06644}}].

\bibitem{Kobayashi:2018bff}
T.~Kobayashi and S.~Tamba, \emph{{Modular forms of finite modular subgroups
  from magnetized D-brane models}},
  \href{https://doi.org/10.1103/PhysRevD.99.046001}{\emph{Phys. Rev. D}
  {\bfseries 99} (2019) 046001}
  [\href{https://arxiv.org/abs/1811.11384}{{\ttfamily 1811.11384}}].

\bibitem{Ohki:2020bpo}
H.~Ohki, S.~Uemura and R.~Watanabe, \emph{{Modular flavor symmetry on a
  magnetized torus}},
  \href{https://doi.org/10.1103/PhysRevD.102.085008}{\emph{Phys. Rev. D}
  {\bfseries 102} (2020) 085008}
  [\href{https://arxiv.org/abs/2003.04174}{{\ttfamily 2003.04174}}].

\bibitem{Kikuchi:2020frp}
S.~Kikuchi, T.~Kobayashi, S.~Takada, T.H.~Tatsuishi and H.~Uchida,
  \emph{{Revisiting modular symmetry in magnetized torus and orbifold
  compactifications}},
  \href{https://doi.org/10.1103/PhysRevD.102.105010}{\emph{Phys. Rev. D}
  {\bfseries 102} (2020) 105010}
  [\href{https://arxiv.org/abs/2005.12642}{{\ttfamily 2005.12642}}].

\bibitem{Kikuchi:2020nxn}
S.~Kikuchi, T.~Kobayashi, H.~Otsuka, S.~Takada and H.~Uchida, \emph{{Modular
  symmetry by orbifolding magnetized $T^2\times T^2$: realization of double
  cover of $\Gamma_N$}},
  \href{https://doi.org/10.1007/JHEP11(2020)101}{\emph{JHEP} {\bfseries 11}
  (2020) 101} [\href{https://arxiv.org/abs/2007.06188}{{\ttfamily
  2007.06188}}].

\bibitem{Kikuchi:2021ogn}
S.~Kikuchi, T.~Kobayashi and H.~Uchida, \emph{{Modular flavor symmetries of
  three-generation modes on magnetized toroidal orbifolds}},
  \href{https://doi.org/10.1103/PhysRevD.104.065008}{\emph{Phys. Rev. D}
  {\bfseries 104} (2021) 065008}
  [\href{https://arxiv.org/abs/2101.00826}{{\ttfamily 2101.00826}}].

\bibitem{Almumin:2021fbk}
Y.~Almumin, M.-C.~Chen, V.~Knapp-P\'erez, S.~Ramos-S\'anchez, M.~Ratz and
  S.~Shukla, \emph{{Metaplectic Flavor Symmetries from Magnetized Tori}},
  \href{https://doi.org/10.1007/JHEP05(2021)078}{\emph{JHEP} {\bfseries 05}
  (2021) 078} [\href{https://arxiv.org/abs/2102.11286}{{\ttfamily
  2102.11286}}].

\bibitem{Kikuchi:2023clx}
S.~Kikuchi, T.~Kobayashi, K.~Nasu, H.~Otsuka, S.~Takada and H.~Uchida,
  \emph{{Remark on modular weights in low-energy effective field theory from
  type II string theory}},
  \href{https://doi.org/10.1007/JHEP04(2023)003}{\emph{JHEP} {\bfseries 04}
  (2023) 003} [\href{https://arxiv.org/abs/2301.10356}{{\ttfamily
  2301.10356}}].

\bibitem{Abe:2013bca}
T.-H.~Abe, Y.~Fujimoto, T.~Kobayashi, T.~Miura, K.~Nishiwaki and M.~Sakamoto,
  \emph{{$Z_N$ twisted orbifold models with magnetic flux}},
  \href{https://doi.org/10.1007/JHEP01(2014)065}{\emph{JHEP} {\bfseries 01}
  (2014) 065} [\href{https://arxiv.org/abs/1309.4925}{{\ttfamily 1309.4925}}].

\bibitem{Abe:2008fi}
H.~Abe, T.~Kobayashi and H.~Ohki, \emph{{Magnetized orbifold models}},
  \href{https://doi.org/10.1088/1126-6708/2008/09/043}{\emph{JHEP} {\bfseries
  09} (2008) 043} [\href{https://arxiv.org/abs/0806.4748}{{\ttfamily
  0806.4748}}].

\bibitem{Kikuchi:2021yog}
S.~Kikuchi, T.~Kobayashi, Y.~Ogawa and H.~Uchida, \emph{{Yukawa textures in
  modular symmetric vacuum of magnetized orbifold models}},
  \href{https://doi.org/10.1093/ptep/ptac035}{\emph{PTEP} {\bfseries 2022}
  (2022) 033B10} [\href{https://arxiv.org/abs/2112.01680}{{\ttfamily
  2112.01680}}].

\bibitem{Dine:1992ya}
M.~Dine, R.G.~Leigh and D.A.~MacIntire, \emph{{Of CP and other gauge symmetries
  in string theory}},
  \href{https://doi.org/10.1103/PhysRevLett.69.2030}{\emph{Phys. Rev. Lett.}
  {\bfseries 69} (1992) 2030}
  [\href{https://arxiv.org/abs/hep-th/9205011}{{\ttfamily hep-th/9205011}}].

\bibitem{Choi:1992xp}
K.-w.~Choi, D.B.~Kaplan and A.E.~Nelson, \emph{{Is CP a gauge symmetry?}},
  \href{https://doi.org/10.1016/0550-3213(93)90082-Z}{\emph{Nucl. Phys. B}
  {\bfseries 391} (1993) 515}
  [\href{https://arxiv.org/abs/hep-ph/9205202}{{\ttfamily hep-ph/9205202}}].

\bibitem{Kobayashi:2020uaj}
T.~Kobayashi and H.~Otsuka, \emph{{Challenge for spontaneous $CP$ violation in
  Type IIB orientifolds with fluxes}},
  \href{https://doi.org/10.1103/PhysRevD.102.026004}{\emph{Phys. Rev. D}
  {\bfseries 102} (2020) 026004}
  [\href{https://arxiv.org/abs/2004.04518}{{\ttfamily 2004.04518}}].

\bibitem{Baur:2019kwi}
A.~Baur, H.P.~Nilles, A.~Trautner and P.K.S.~Vaudrevange, \emph{{Unification of
  Flavor, CP, and Modular Symmetries}},
  \href{https://doi.org/10.1016/j.physletb.2019.03.066}{\emph{Phys. Lett. B}
  {\bfseries 795} (2019) 7} [\href{https://arxiv.org/abs/1901.03251}{{\ttfamily
  1901.03251}}].

\bibitem{Novichkov:2019sqv}
P.P.~Novichkov, J.T.~Penedo, S.T.~Petcov and A.V.~Titov, \emph{{Generalised CP
  Symmetry in Modular-Invariant Models of Flavour}},
  \href{https://doi.org/10.1007/JHEP07(2019)165}{\emph{JHEP} {\bfseries 07}
  (2019) 165} [\href{https://arxiv.org/abs/1905.11970}{{\ttfamily
  1905.11970}}].

\bibitem{Baur:2019iai}
A.~Baur, H.P.~Nilles, A.~Trautner and P.K.S.~Vaudrevange, \emph{{A String
  Theory of Flavor and $\mathscr {CP}$}},
  \href{https://doi.org/10.1016/j.nuclphysb.2019.114737}{\emph{Nucl. Phys. B}
  {\bfseries 947} (2019) 114737}
  [\href{https://arxiv.org/abs/1908.00805}{{\ttfamily 1908.00805}}].

\bibitem{Abe:2009vi}
H.~Abe, K.-S.~Choi, T.~Kobayashi and H.~Ohki, \emph{{Non-Abelian Discrete
  Flavor Symmetries from Magnetized/Intersecting Brane Models}},
  \href{https://doi.org/10.1016/j.nuclphysb.2009.05.024}{\emph{Nucl. Phys. B}
  {\bfseries 820} (2009) 317}
  [\href{https://arxiv.org/abs/0904.2631}{{\ttfamily 0904.2631}}].

\bibitem{Betzler:2019kon}
P.~Betzler and E.~Plauschinn, \emph{{Type IIB flux vacua and tadpole
  cancellation}}, \href{https://doi.org/10.1002/prop.201900065}{\emph{Fortsch.
  Phys.} {\bfseries 67} (2019) 1900065}
  [\href{https://arxiv.org/abs/1905.08823}{{\ttfamily 1905.08823}}].

\end{thebibliography}\endgroup
\bibliographystyle{JHEP} 

\end{document}